\documentclass[structabstract]{aa}
\usepackage{graphicx}% figures
\usepackage{txfonts} % roman
\usepackage{natbib}
\bibpunct{(}{)}{;}{a}{}{,} %  A&A style for bibliography

\usepackage{lscape}    
\renewcommand\footnoterule{  \kern -3pt  \hrule \kern 2pt }  

\begin{document}

\title{Ionization structure and chemical abundances of the Wolf$-$Rayet nebula NGC\,6888 with integral field spectroscopy\thanks{Based on observations collected at the Centro Astron\'omico Hispano Alem\'an (CAHA) at Calar Alto, operated jointly by the Max$-$Planck$-$Institut f\"ur Astronomie and the Instituto de Astrof\'isica de Andaluc\'ia (CSIC).}}

%\subtitle{}
\author{A. Fern\'andez$-$Mart\'in \inst{1} \thanks{e$-$mail: alba@iaa.es} ,
							 D. Mart\'in$-$Gord\'on \inst{1},
							 J.M. V\'ilchez \inst{1} , 
							  E. P\'erez Montero  \inst{1},
						     A. Riera \inst{2}
							 \and  S. F. S\'anchez  \inst{3}
				}
\institute{Instituto de Astrof\'isica de Andaluc\'ia (IAA$-$CSIC),  Glorieta de la Astronom\'ia  S/N, 18008 Granada, Spain 
						    \and Departament de F\'isica i Enginyeria Nuclear, EUETIB, Universitat Polit\'ecnica de Catalunya, C. Comte Urgell 187, 08036, Barcelona, Spain
						    \and Centro Astron\'omico Hispano Alem\'an, Calar Alto, CSIC$-$MPG, C/Jesús Durb\'an Remón 2$-$2, E$-$04004 Almer\'ia, Spain
				}                                                                       
%email: alba@iaa.es

% \date{Received.....; accepted..... } %set by editors and publisher

% 5 {} token are mandatory

\abstract % ORDEN:{context}{aims}{methods}{results}{discussion}
{The study of nebulae around Wolf$-$Rayet (WR) stars gives us clues about the mass$-$loss history of massive stars, as well as about the chemical enrichment of the interstellar medium (ISM).}
{This work aims to search for the observational footprints of the interactions between the ISM and stellar winds in the WR nebula NGC\,6888 in order to understand its ionization structure, chemical composition, and kinematics.}
{We have collected a set of integral field spectroscopy observations across NGC\,6888, obtained with PPAK in the optical range performing both 2D and 1D analyses. Attending to the 2D analysis in the northeast part of NGC\,6888, we have generated maps of the extinction structure and electron density. We produced statistical frequency distributions of the radial velocity and diagnostic diagrams. Furthermore, we performed a thorough study of integrated spectra in nine regions over the whole nebula.}
{The 2D study has revealed two main behaviours. We have found that the spectra of a localized region to the southwest of this pointing can be represented well by shock models assuming $n=1000~ cm^{-3}$, twice solar abundances, and shock velocities from 250 to 400 km~s$^{-1}$. With the 1D analysis we derived electron densities ranging from $<$100 to 360 cm$^{-3}$. The electron temperature varies from $\sim$7700~K to $\sim$10\,200~K. A strong variation of up to a factor 10 between different regions in the nitrogen abundance has been found: N/H appears lower than the solar abundance in those positions observed at the edges and very enhanced in the observed inner parts. Oxygen appears slightly underabundant with respect to solar value, whereas the helium abundance is found to be above it.\\
We propose a scenario for the evolution of NGC\,6888 to explain the features observed. This scheme consists of a structure of multiple shells: i) an inner and broken shell with material from the interaction between the supergiant and WR shells, presenting an overabundance in N/H and a slight underabundance in O/H; ii) an outer shell very intense in [OIII]$\lambda$5007\AA\ corresponding to the main sequence bubble broken up as a consequence of the collision between supergiant and WR shells. Nitrogen and oxygen do not appear enhanced here, but helium appears enriched. iii) And finally it includes an external and faint shell that surrounds the whole nebula like a thin skin representing the early interaction between the winds from the main sequence star with the ISM for which typical circumstellar abundances are derived.}
{}

\titlerunning{Study of NGC 6888 with PPAK}   
\authorrunning{Fern\'andez$-$Mart\'in et al.}  
\maketitle

%________________________________________________________________

\section{Introduction\label{intro}}
Wolf$-$Rayet (WR) stars are the evolved descendants of massive OB stars ($M>25M_{\sun}$), and they are believed to constitute the previous phase in their evolution before supernova explosion. Many of them have been shown to be surrounded by nebular emission in a structure of shells (ring nebula). \citet{1965ApJ...142.1033J} were the first to suggest that these nebulae were due to material ejected from the star sweeping up the surrounding interstellar medium (ISM). \citet{1969SvA....12..757P} and \citet{1977ApJ...218..377W} both made models deriving analytical solutions about dynamic evolution of shock bubbles created by interaction between the ISM and the stellar wind. One of the most comprehensive models describing the basic optical morphology of bubbles in WR stars was produced by \citet{1996A&A...316..133G}. Their hydrodynamical simulations take the evolution and mass$-$loss history into account over the lifetime of the central star from the main sequence (MS) to the WR phase. In summary, stellar winds from the MS star sweep up the ISM, forming a hot bubble surrounded by a thin shell; later, dense material from the red supergiant (RSG) winds partially fills up the bubble creating another internal shell; finally, in the WR stage, the star develops another mass$-$loss episode with a fast wind that reaches the RSG shell and can eventually break up the MS bubble, creating a wind$-$blown bubble (WBB).\\

NGC\,6888 (S105) is a proto$-$typical WBB associated with the WN6 star WR\,136 (HD 192163) \citep{1965ApJ...142.1033J}. It was first reported by \citet{1959ApJS....4..257S}, and its relative proximity and large optical angular size has made it one of the best$-$studied examples of this class of objects, which are observed in a wide wavelength range. The basic parameters of star and nebula are summarized in Table \ref{table:parameter}.

\begin{table*}
		 \caption{Physical properties of WR\,136 and NGC\,6888.}    % title of Table
		\label{table:parameter}     % is used to refer this table in the text
		\centering                          % used for centering table
		\begin{tabular}{ l l l l l}	\hline
				\\
             Object &  Parameter   &  Value   & Reference   \\  \hline \hline
			WR\,136	& ($\alpha$,$\delta$) (J2000) & (20:12:06.55,+38:21:17.8) & \citet{2001NewAR..45..135V}   \\ 	
  				             & Spectral type & WN6& \citet{2001NewAR..45..135V}   \\ 	
				& $M_{\mathrm{\star}} $($M_{\mathrm{\sun}}$) &  15   & Hamann et al. (2006) \\
			   &  Distance (kpc )& 1.26  & \citet{2001NewAR..45..135V} \\
				& $R_{\mathrm{G}} $ (kpc )&10 & Esteban et al. (1992)  \\
		%	   &  v (mag) & 7.65 &  \citet{2001NewAR..45..135V} \\
 				& $M_{\mathrm{v}} $ (mag) & -4.69 & Hamann et al. (2006) \\
 			  & $E_{\mathrm{b-v}} $ (mag) & 0.45 & Hamann et al. (2006)  \\ 		
		\\		
			NGC\,6888  &   Angular size (arcmin$^{2}$)& 12x18 & \citet{1983ApJS...53..937C}  \\
				& $V_{\mathrm{exp}}$  (km~s$^{-1}$)  & 55$-$110 &  \citet{1970SvA....14...98L} \\
				&$ M_{\mathrm{ionized}}$ ($M_{\mathrm{\sun}}$) & 5   &  Wendker et al. (1975) \\
				& $M_{\mathrm{neutral}}$ ($M_{\mathrm{\sun}}$)& 40 &  \citet{1988MNRAS.235..391M}  \\
				\hline		\\
			\end{tabular}
		\end{table*}

The ellipsoidal shell appears to be geometrically prolate and to have a highly filamentary structure \citep{1970SvA....14...98L}. The expansion velocity of the shell varies in the range 55$-$110 km~s$^{-1}$ \citep[][among others]{1970SvA....14...98L,1988MNRAS.235..391M,1992ApJ...390..536E}. Spectroscopy of the nebular shell shows that the ionized gas is enriched with nitrogen and helium \citep{1975A&A....42..173W} and is slightly underabundant in oxygen \citep{1981ApJ...245..154K,1992ApJ...390..536E}. This could be explained as a consequence of the mixing of interstellar and stellar wind material \citep{1981ApJ...245..154K} and the transport from the core of the star to the shell during the RSG phase \citep{1992ApJ...390..536E}. \citet{1995AJ....109.2257M} presented IRAS images obtaining two concentric shells associated with WR\,136. With the inferred masses, he proposed that material of the inner shell comes from the mass lost by the star in the RSG phase and the outer shell formed from the early O star phase. \citet{2000AJ....120.2670G} observed H${\alpha}$ and [OIII]$\lambda$5007\AA{} emission from NGC\,6888, finding an offset between both images and suggest that these offsets are due to the shock from the WR bubble expanding into the circumstellar medium. This contrast between the H${\alpha}$ and [OIII]$\lambda$5007\AA{} morphology is shown better in the images from the WFPC2 on the HST \citep{2000AJ....119.2991M}.\\

Bubbles blown by massive stars are filled by the fast stellar wind that interacts strongly with the ISM, and some fraction of the kinetic energy is converted into thermal energy heating the gas at X$-$ray emitting temperatures \citep{1977ApJ...218..377W}. NGC\,6888 was the first WBB that has been detected in X$-$rays, it was made with \textit{Einstein} by \citet{1988Natur.332..518B}. The \citet{1994A&A...286..219W} X$-$ray observation did not detect a filled bubble emitting as models predicted, but a filamentary X$-$ray structure was instead found following the H${\alpha}$ features of the nebula. In addition, spectra in the shell were fit to a double$-$temperature plasma model with a high$-$temperature component at $ T\sim8.5\times10^{6}$ K and a low$-$temperature component at $T\sim1.3\times10^{6}$ K without detecting significant temperature variation between different areas of the nebula \citep{2005ApJ...633..248W}. In the last X$-$ray study of this nebula, \citet{2011ApJ...728..135Z} have found a bi$-$modal distribution of the X$-$ray emitting plasma, and made models to explain how the soft$-$X rays come from evaporation of clumps. They also discuss whether the heating conduction is efficient in the hot bubble.

NGC\,6888 represents one of the most suitable candidates for searching for the observational footprint of the interactions between ISM and stellar winds. In addition, it can give us clues about the mass$-$loss history of massive stars, as well as about the enrichment of the ISM \citep{1992A&A...259..629E,2010A&A...517A..27M}. Despite all the studies already done on NGC\,6888, the spatial extent of the chemical enrichment and the homogeneity of the physical properties and structure are still unknown , since no 2D measures of their spectroscopic properties (and imaging, simultaneously) have been done so far. To study the morphology and chemodynamics in 2D, we decided to include NGC\,6888 as a target in our programme of integral field spectroscopy (IFS) to perform the spatial analysis of spectral maps and compare them with 1D studies of selected integrated spectra.\\

The paper is organized as follows. The observations and data reduction are described in Sect. \ref{obsandred}. Description, analysis, and results of the physical conditions of the 2D study are presented in Sect. \ref{2dim}, and Sect. \ref{1dim} is devoted to study selected 1D spectra. The discussion and the summary of the main conclusions are given in Sects. \ref{discussion} and \ref{conclusions}, respectively .

%________________________________________________________________

\section{Observations and data reduction\label{obsandred}}		
\subsection{Observations \label{obs}}
Observations were carried out in 2005 July at the 3.5$-$m telescope of the Centro Astron\'omico Hispano Alem\'an (CAHA) at the observatory of Calar Alto (Almería, Spain) with the Potsdam Multi$-$Aperture Spectrograph instrument (PMAS) \citep{2005PASP..117..620R}  in PPAK mode \citep{2006PASP..118..129K}. The PPAK configuration consists of 331 science fibres with a diameter of 2.7 arcsec covering a hexagonal field of view (FoV) of 74$\times$65 arcsec$^{2}$. The surrounding sky is sampled by 36 additional fibres distributed in six bundles located following a circle at about 90\arcsec \,from the centre. Furthermore, there are 15  fibres for calibration purposes \citep[see Fig. 5 in][]{2006PASP..118..129K}.

\begin{figure*}
  \includegraphics[width=\textwidth]{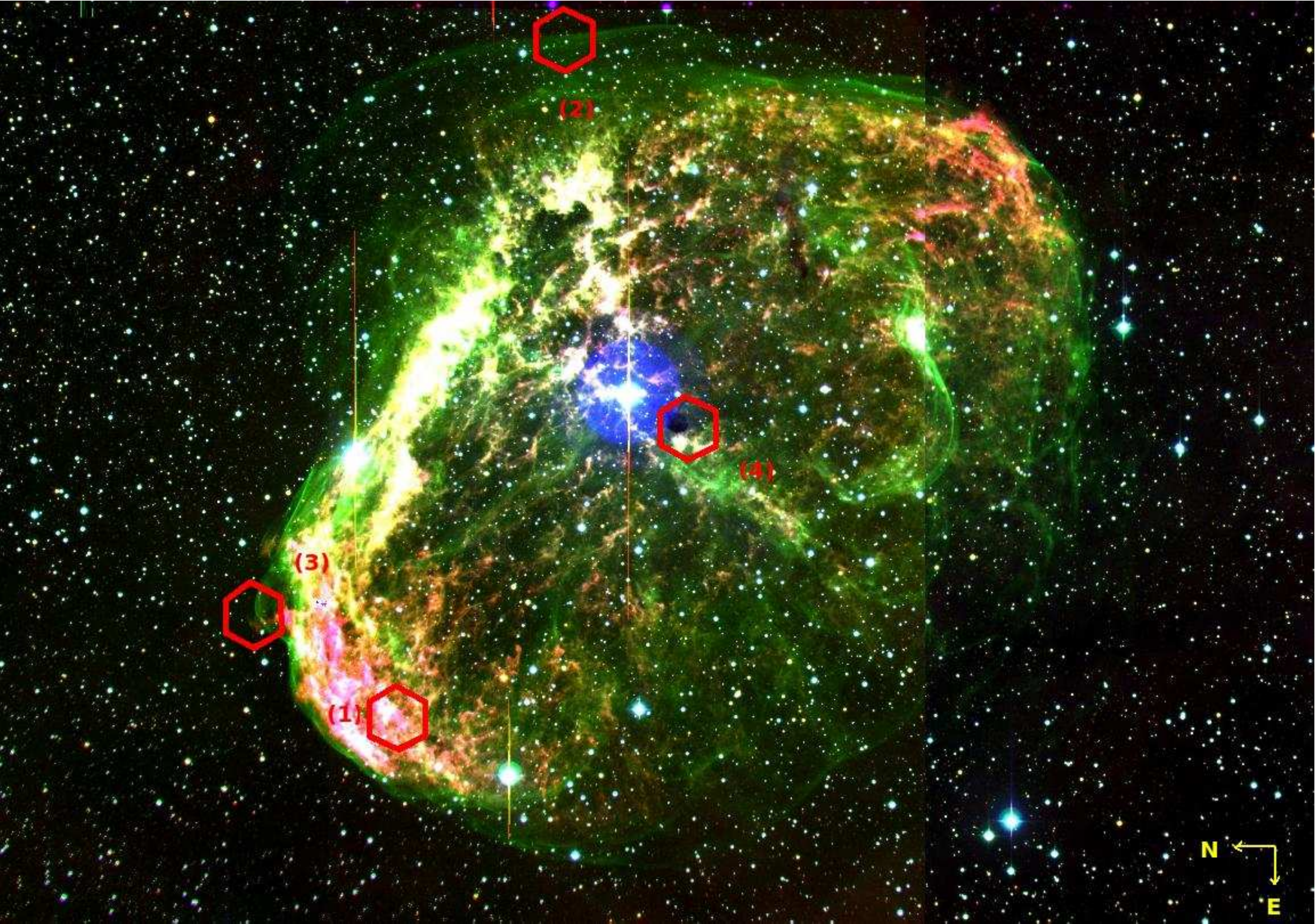}
		\caption{RGB composited image of  NGC\,6888 taken with the WFC at the Isaac Newton Telescope (INT). Red corresponds to the H${\alpha}$ emission,  green to [OIII], and blue to [OII]. [OII] emission is very faint but noticeable as the pink shade in the mix of H${\alpha}$ and [OII].  North is left and east at the bottom. Red hexagons show the zones of our IFS observations: (1) X$-$ray, (2) edge, (3) mini$-$bubble, and (4) bullet.}
       \label{fig:zone}
		\end{figure*}

We chose the zones of NGC\,6888 to be mapped based mainly on the narrow band images observed by our group in the Isaac Newton Telescope (INT) with the Wide Field Camera (WFC). Figure \ref{fig:zone} shows an RGB composited image of the nebula; H${\alpha}$ and [OIII] emissions represented by red and green colours respectively, the [OII] emission, in blue, is fainter but noticeable as the pink shade in the mix with H${\alpha}$. Here, we present IFS observations of NGC\,6888 in four zones of the nebula (see Fig. \ref{fig:zone}): the first zone (1) in the  northeast area,  was chosen because it can be a good candidate for presenting signs of shocked gas, and it is known that this area emits in X$-$rays. Two other pointings, (2) and (3), were chosen in the outer zones of NGC\,6888 where H${\alpha}$ emission is very weak but an [OIII] $“skin”$ is observed in Fig. \ref{fig:zone}. Finally, another zone (4) was selected to study the centre of the nebula where we can observe a striking dark region in all the wavelengths or a knot that could be associated to our object. We have called all these regions : X$-$ray zone (1), edge (2), mini$-$bubble(3), and bullet (4) zones, as marked in  Fig.\ref{fig:zone}.

We used three different gratings to obtain information for the most important emission lines in the optical spectral range. The R1200 grating used was centred in two wavelengths covering the spectral range from 6220 to 6870 \AA{} and from 4320 to 5060 \AA{} with a dispersion of  0.3\AA{}/pix, giving a spectral resolution of $\sim$2.4 \AA{}FWHM at 6300\AA{}. A V1200 grating with a dispersion of  0.35\AA{}/pix was used that covered the spectral range from 3660 to 4430 \AA{} with a spectral resolution of FWHM$\sim$2.7\AA{} at 4047\AA{}. This setup overlaps 100 \AA{} with the  wavelength range of the R1200 blue grating. Finally, we used a low$-$resolution configuration with the V300 grating, covering the spectral range from 3660 to 7040 \AA{} with a dispersion of 1.67\AA{}/pix, giving a spectral resolution of $\sim$8.7 \AA{}FWHM at 5577 \AA{}. Table  \ref{table:log} shows the observational log for NGC\,6888 in the four regions and three exposures were taken in each pointing. Moreover, in the X$-$ray zone with R1200 gratings,  we adopted a dithering scheme with three positions having offsets measured from the centre of $\Delta\alpha=-1.35\arcsec$, $\Delta\delta=-0.78\arcsec$ for the second pointing and $\Delta\alpha=-1.35\arcsec$, $\Delta\delta=+0.78 \arcsec$ for the third. This pattern allowed us to fill up the gap between fibres.

The weather was photometric throughout the observations with the typical seeing subarcsecond. Two spectrophotometric standard stars, Hz\,44 and BD+28\,4211, were observed for all the setups. The necessary bias frames, continuum, and arcs were acquired.

\begin{table*}%[h!]
		 \caption{NGC\,6888 PPAK observational log.}   
		\label{table:log}    
		\centering                       
		\begin{tabular}{l c c c c c c c}
		\hline
		Zone &  Coordinates (J2000) &Grating & Spectral range & Exp. time & Airmass  & Date &Comments \\
    	        & ($\alpha$,$\delta$)  & & ( \AA{} ) &  (s) & &\\
		\hline\hline
        X$-$ray zone (1) &  (20:12:39.8 , +38:25:34.5) & V300	& 3640$-$7040 & 1350 & 1.00 & July, 5& \\
		&  & R1200 & 6200$-$6870 & 1440 (4320$^{\dagger}$)& 1.02 & July, 1 & dithering ($\times$3)\\
		 & & R1200 & 4320$-$5060 & 1620 (4860$^{\dagger}$)& 1.03 & July, 2 &  dithering ($\times$3) \\
		 & & V1200 & 3660$-$4430 & 1620 (4860$^{\dagger}$)& 1.02 &July, 4 & dithering ($\times$3)$^{a}$ \\		
		Edge  (2) & (20:11:30.7 , +38:22:38.2) &  V300 & 3640$-$7040 & 1350 & 1.01 &July, 5 &\\
         & &R1200 & 6200$-$6870 & 1620 & 1.00 & July, 1 &\\
		 & &  R1200 & 4320$-$5060 &  1620 & 1.00 &July, 2 &\\	
		Mini$-$bubble (3) & (20:12:29.0 ,+38:28:39.0) & V300 & 3640$-$7040 & 1350 & 1.00 &July, 6 &\\
		Bullet  (4) & (20:12:09.6 ,+38:20:14.9) & V300 & 3640$-$7040 & 1350 & 1.00 &July, 6 &\\
		\hline
		\multicolumn{3}{l}{$^{\dagger}$ Total time including the three dithered exposures.}\\
		\multicolumn{8}{l}{$^{a}$ One of these dithering positions suffers from dome$-$lamp contamination. Only two of the three exposures are useful.}\\
		\end{tabular}
		\end{table*}

					\subsection{Data reduction \label{red}}
The data were reduced using the R3D software \citep{2006AN....327..850S}  in combination with the Euro3D packages \citep{2004AN....325..167S} and IRAF \footnote{The Image Reduction and Analysis Facility IRAF is distributed by the National Optical Astronomy Observatories, which are operated by Association of Universities for Research in Astronomy, Inc., under cooperative agreement with the National Science Foundation.}. The reduction consisted of the standard steps for fibre$-$fed IFS observations.

A master bias frame for each night of observations was created by averaging the corresponding bias frames observed during the night and subtracted from all the images. The different exposures taken at the same position on the sky were combined to reject the cosmic rays using the package IMCOMBINE of IRAF.

Using continuum$-$illuminated exposures taken before the science exposures, we determined the location of spectra on the detector for each pixel along the dispersion axis.  Then, each spectrum was extracted from the science and standard star frames, co$-$adding the flux within an aperture on the location of the spectral peak in the raw data using the tracing information, and  storing in a 2D image called row$-$stacked$-$spectrum (RSS) \citep{2004AN....325..167S}. Furthermore, we checked that the contamination from flux coming from adjacent fibres using this aperture was negligible \citep{2004PASP..116..565B,2006AN....327..850S} . For a given aperture and $FWHM\sim(0.5\times~aperture)$ we found a level of cross$-$talk that is always $<$10$\%$. This seems to be an acceptable compromise between maximizing the recovered flux and minimizing the cross$-$talk.

Wavelength calibration was performed in a two$-$step procedure using arc calibration$-$lamp exposures (a neon lamp for the R1200 grating in the range from 6220 to 6870 \AA{} and He for the rest of the pointings). In the first step, distortion correction was carried out to avoid the curvature. Then, the wavelength coordinate system was determined by identifying the lines. Finally, the distortion and dispersion solutions were applied over the science data. The accuracy achieved was better than $\sim$0.1\AA{}(rms) for the arc exposures in every spectral range. We performed a further wavelength \textquotedblleft re$-$centring\textquotedblright using sky emission lines present in the observed wavelength ranges.

Corrections to minimize the differences between fibre$-$to$-$fibre transmission throughput were also applied, creating a fibre$-$flat with an exposure of a continuum. All the science frames were divided by the obtained fibre$-$flat.

Observations of spectrophotometric standard stars were used to perform the flux calibration. Only Hz\,44 was observed all the nights, so we used it for our primary calibration. With the E3D software a spectrum for the calibration star was created each night by co$-$adding the central fibres of the standard star frames. Then, with IRAF tasks, a sensitivity function was determined. Subsequently, we checked that the flux$-$calibration for the two standard stars was self$-$consistent  for all the nights observed and with the library values. Finally, flux calibration was applied to all the science frames.

To achieve an accurate sky subtraction is an important issue in data reduction.  Although PPAK has additional fibres for sampling the sky, our object is very extended, and all the sky fibres are located within an area containing signals from the nebula. Extra frames to sample the background near NGC\,6888 were not available. For each night, we adopted different sky$-$subtraction schemes. For the night when we used the V1200 grating, an external sky$-$frame was observed. Since the sky and object exposures were taken within a few minutes of each other in the same night and under similar weather conditions, we decided to estimate a sky spectrum using this extra frame. To do so, the spectra of all individual fibres of the sky were combined with a mean in a single spectrum and a 2D spectrum was created with the value in each fibre of the combined  spectrum. This sky was subsequently subtracted from every science spectrum. This method was used in observations with the high$-$resolution gratings (R1200 and V1200), and no satisfactory sky frame was available for the night of V300 observations.

Considering the short wavelength range and the airmass of the observations, and using \citet{1982PASP...94..715F}, we estimated that the offsets due to the differential atmospheric refraction (DAR) were always smaller than one third of the diameter fibre. Correction for DAR was not necessary in our data.

%________________________________________________________________

\section{Two$-$dimensional study\label{2dim}}
Once data had been reduced, we first carried out a basic 2D study to describe the observed morphology of the four pointings in several emission lines. Later, in zones with a dithering scheme, we performed a more detailed analysis.
				\subsection{Morphology in the emission lines \label{morphology}}
\begin{figure*}
		  \centering
		 \includegraphics[width=17cm]{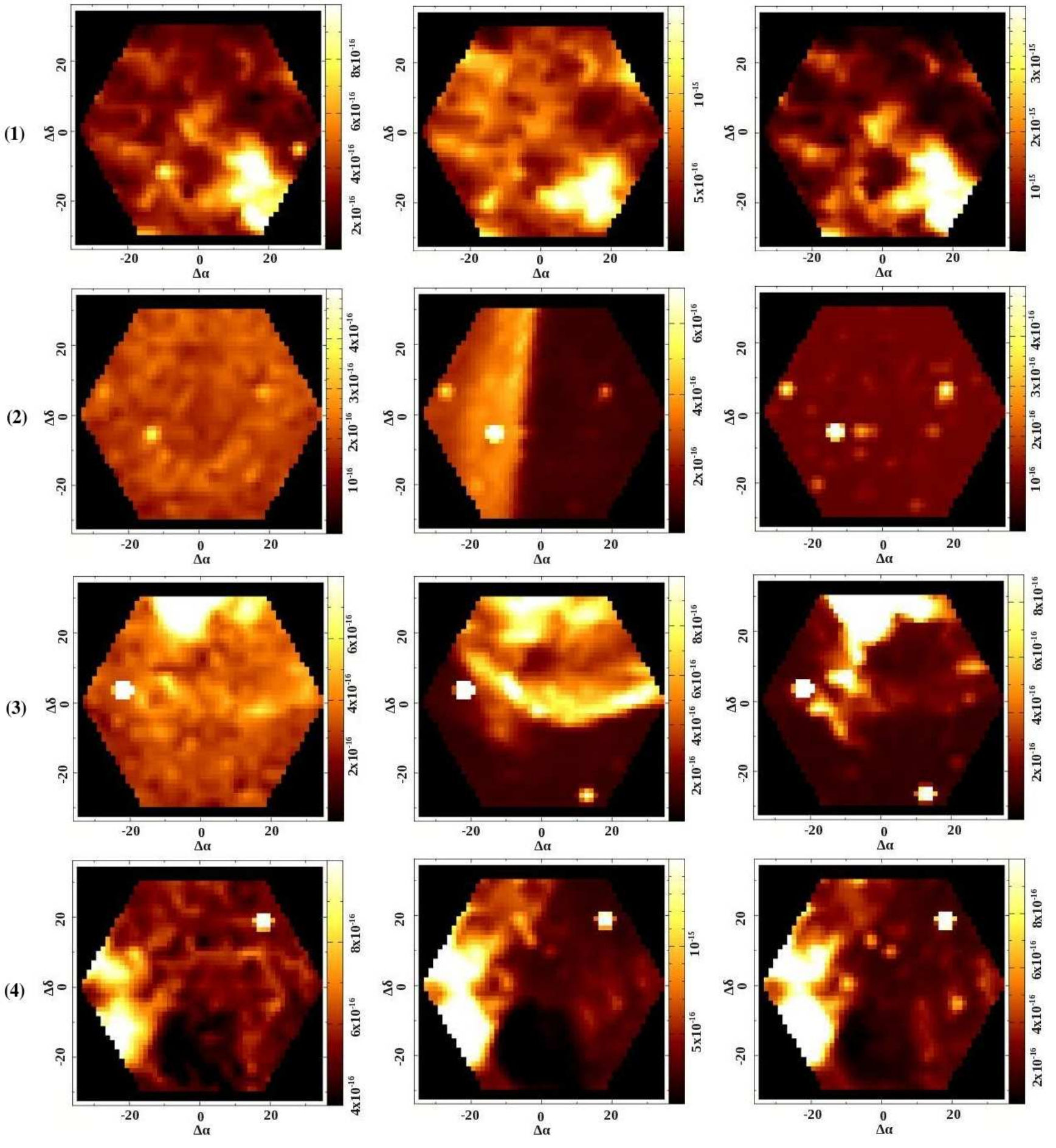}
	\caption{Images of the four pointings of NGC\,6888. Columns are emission lines: from left to right [OII]$\lambda\lambda$3726,3728\AA{}, [OIII]$\lambda$5007\AA{}, and [NII]$\lambda\lambda$6548,6584\AA{}+H${\alpha}$. The four different rows show the four regions observed; from top to bottom: (1) X$-$ray zone, (2) edge, (3) mini$-$bubble, and (4) bullet. In all the maps, north is down and east to the left.}
     \label{fig:morphology_all}
		\end{figure*}

Using E3D software, we generated interpolated images of the four observed regions in three  wavelength ranges including the lines: [OII]$\lambda\lambda$3726,3728\AA{}, [OIII]$\lambda$5007\AA{}, and [NII]$\lambda\lambda$6548,6584\AA{}+H${\alpha}$.  Figure \ref{fig:morphology_all} displays these images. It can be seen two different patterns in all the regions, [OII] and H${\alpha}$+[NII] present similar behaviour, while [OIII] shows a different structure. This can be understood as the signature of the ionization structure of the nebula.

In the X$-$ray emitting region, most of the emission is located in the northwest and seems spread towards the northeast and centre. The high ionization emission (i.e. [OIII]) shows two main peaks, one in the northwest and the other in the southwest.

%\clearpage

The second zone (Edge) was chosen to be observed because is placed in a big bubble of [OIII] that appears surrounding all the nebula. It can be observed in Fig. \ref{fig:morphology_all}, that in the [OIII]$\lambda$5007\AA{} emission$-$line, a discontinuity separates the nebula into two different regions, while it appears uniform in the low ionization emission$-$line images.

The little bubble in the northern region  of the nebula (Mini$-$bubble) is cleary seen in the [OIII] images, but appears very diffuse in H${\alpha}$ in Fig. \ref{fig:zone}. In the PPAK observations of this region different features can be distinguished, a bright zone to the south in all the wavelengths and a semi$-$circle structure only seen in the [OIII] image.

Finally, the last row of Fig. \ref{fig:morphology_all}  corresponds to a darker region placed near the centre of NGC\,6888 that  could be a sort of \textquotedblleft bullet\textquotedblright \,crossing the nebula. This region appears quite similar in the three wavelength maps. The dark zone in the north is very faint in all the emission lines, though again low ionization lines show slight differences from [OIII].\\

				\subsection{Data$-$cube analysis of the X$-$ray zone \label{cubes}}
The bi$-$dimensional structure of the three dithering observations was studied in detail by means of the creation of data cubes. We performed this kind of analysis only for the X$-$ray zone as observed with R1200 gratings in two wavelength ranges because for the other pointings dithering was not available. In the next sections (\ref{extinc}, \ref{diagnostic}, and \ref{kine}), all the results refer to the X$-$ray zone.

As explained above, the 331 science fibres lead the light from different positions of the target to the pseudo$-$slit. When dithering exposures (to fill holes) were performed, it is important to combine them when knowing the overlapping areas and registering the spectra to their positions in the sky. With this aim, we used the R3D software to construct a grid with a pixel scale about one third of the original fibres ($\sim$1 arcsec), obtaining easier$-$to$-$use data cubes with two spatial and one spectral dimension. Therefore, two cubes of the X$-$ray region in the wavelength ranges 6220$-$6870 \AA{} and 4320$-$5060 \AA{} were built.\\

One of our objectives was to obtain maps from the datacubes of the main emission lines, which could then be analysed to describe the 2D ionization structure and physical properties of the X$-$ray emitting region of NGC\,6888. This procedure was performed using our own routine, which fits each line with a Gaussian function. We introduce four parameters as a reference to the fitting (the central wavelength, the FWHM, the line wavelength range, and the corresponding continuum range) and maps are returned with useful information in each pixel for each selected line (i.e. flux, FWHM, central wavelength, error of flux, continuum, etc.).

As a first step, we fit all the lines by a single Gaussian for the two data cubes obtained. For the grating placed in the red range (from 6200 to 6870 \AA{}), seven lines were fit: [SIII]$\lambda$ 6312\AA{}, H${\alpha}$, [NII]$\lambda \lambda$6548,6584\AA{}, HeI$\lambda$6678\AA{}, and [SII]$\lambda \lambda$6717,6731$\AA{}$.  In the other spectral range (from 4320 to 5060 \AA{}), another seven lines were fit: H${\gamma}$, H${\beta}$, [OIII]$\lambda \lambda$4363,4959,5007\AA{}, and HeI$\lambda \lambda$4471,5015\AA{}. In addition, to prevent contamination by low signal$-$to$-$noise (S/N) data, we masked all pixels with an S/N lower than 5 out. Under these conditions, maps were used in the analysis that we present in the next sections.

Two main problems appeared in the maps creation. First, with those fainter lines placed in zones with high noise, the fits of the automatic routine were not good in all the spaxels, and some adjacent signal was added to the fit, giving a wrong measurement of the line flux. It happened with three lines:  [SIII]$\lambda$6312$\AA{}$, HeI$\lambda$4471\AA{}, and HeI$\lambda$5015\AA{}. With the obtained maps for these lines, masks were created in order to not take bad pixels into account in the analysis. The second problem that we found was fitting the [OIII]$\lambda$4363\AA{}. This line is located very close to the strong HgI$\lambda$4358\AA{} night sky line, in particular, both lines are overlapped in our object, and we cannot resolved them, so we could not obtain workable maps of [OIII]$\lambda$4363\AA{} for this nebula. Without the measure of this line, it was not possible to obtain the electron temperature map. Furthermore, no other auroral line, like [NII]$\lambda$5755\AA{}, was measured in the cubes with a high S/N (but see Sect. \ref{int_prop}).

\subsubsection{Extinction and electron density maps \label{extinc}}
The reddening coefficient c(H${\beta}$) was derived from the H${\alpha}$/H${\beta}$ line ratio. We used an intrinsic Balmer emission line ratio of $H{\alpha}/H{\beta}=2.81$ obtained from  the public software of \citet{1995MNRAS.272...41S} assuming Case B recombination with an electron temperature of $T_{\mathrm{e}}\sim10^{4}$ K and an electron density of  $n_{\mathrm{e}}\sim10^{2}$ cm$^{-3}$.

\begin{figure}
		  \centering
 \includegraphics[width=7cm]{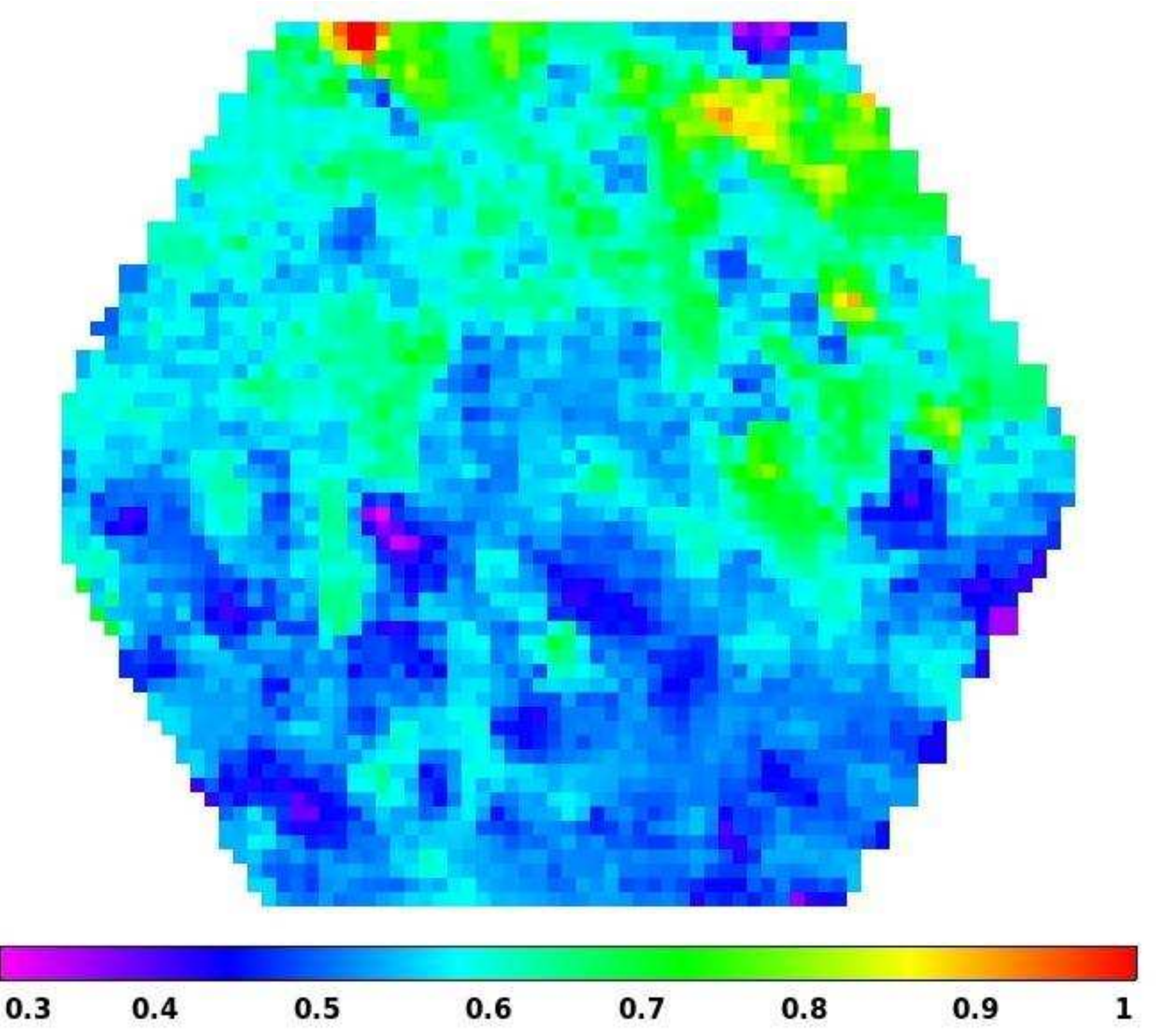}
		\caption{Reddening coefficient map, c(H${\beta}$), derived from the H${\alpha}$/H${\beta}$ line ratio. Orientation as in Fig. \ref{fig:morphology_all}.}
    \label{fig:reddening}
		\end{figure}

Although other Balmer lines were detected in our wavelength spectral range, H${\alpha}$ and H${\beta}$ were the lines with the highest S/N. We performed the estimation of c(H${\beta}$) including  H${\gamma}$. The average value obtained was similar but the errors increased, and the S/N in the H${\gamma}$ flux was lower than 5 in 10$\%$ of the spaxels, so finally we only used H${\alpha}$ and H${\beta}$. However, we checked that in the spaxels with good S/N in H${\gamma}$ both derivations were consistent. The derived c(H${\beta}$) map is shown in Fig. \ref{fig:reddening}, where a non$-$uniform, rather filamentary structure can be clearly distinguished, with values ranging from $c(H{\beta})=0.3$ to 1.00, with a mean value of 0.57. Adopting $E(B-V)=0.692 \times c(H{\beta})$ and using the \citet{1989ApJ...345..245C} extinction law with $R_{\mathrm{v}}=3.1$, E(B-V) was computed with values in the range 0.21$-$0.88. In general terms, the structure presented in this map is in good agreement with values reported in the literature. Using the flux density at 21cm and the flux in H${\alpha}$, \citet{1975A&A....42..173W} estimated an absorption over the whole nebula of $A_{\mathrm{V}} \sim2.05$ mag ($E(B-V) =A_{\mathrm{V}}/3.1=0.66$). \citet{1992ApJ...390..536E}  obtained values of the c(H${\beta}$) reddening coefficient of 0.28 for the blueshifted component and 0.50 for the redshifted one, correspond to E(B-V) of 0.19 and 0.34, respectively. Some of these values are slightly lower than ours, but their slit positions do no spatially coincide with our observed zones.

Finally, all the emission$-$line maps were reddening$-$corrected using the derived c(H${\beta}$) map. Observing their appearance we checked the morphology explained in Sect. \ref{morphology} but now with more lines: H${\alpha}$, H${\beta}$, [SII], [NII], and He I with similar patterns as examples of low ionization and the doublet of [OIII] as  high ionization.\\

\begin{figure}
	  \centering
		 \includegraphics[width=7cm]{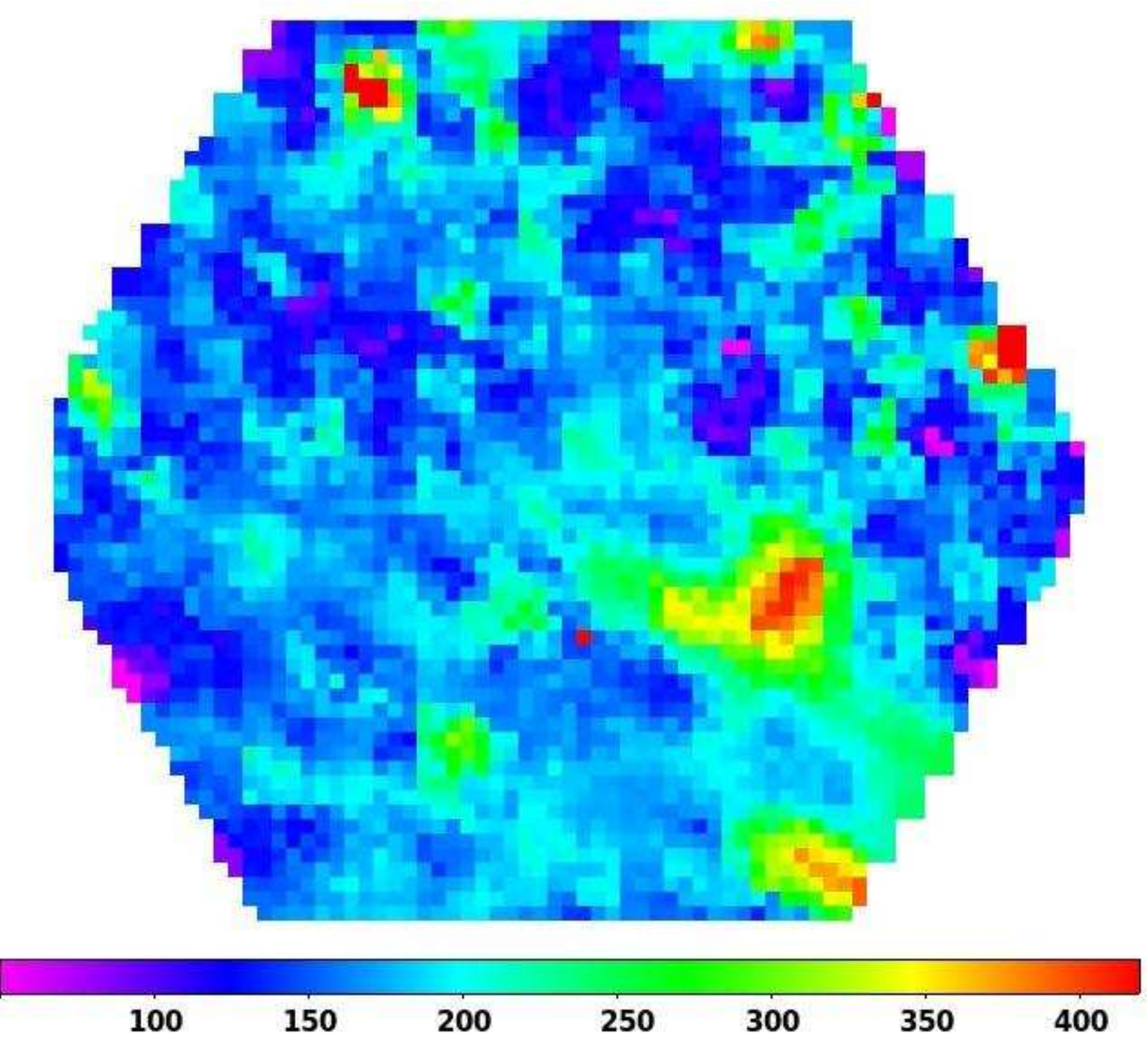}
		\caption{Electron density n$_{\mathrm{e}}$ map derived from the [SII]$\lambda\lambda$6717/6731 line ratio. The n$_{e}$ distribution ranges from $\sim$50 cm$^{-3}$ to a maximum peak of 410 cm$^{-3}$, with a mean of 178 cm$^{-3}$. Orientation as in Fig. \ref{fig:morphology_all}.}
        \label{fig:density}
  		\end{figure}

The electron density (n$_{\mathrm{e}}$) map of the X$-$ray region was produced from the [SII]$\lambda\lambda$6717/6731 ratio using the package TEMDEN of IRAF based in a five$-$level statistical equilibrium model \citep{1987JRASC..81..195D,1995PASP..107..896S}. The spatial distribution of the electron density can be seen in Fig. \ref{fig:density}. The map shows a rather filamentary aspect that is more or less uniform with a mean value of 178 cm$^{-3}$. The main significant feature in this spatial distribution is the peak density ($\sim$400 cm$^{-3}$) coincident with the maximum flux emission of all the species detected. Our highest n$_{\mathrm{e}}$ values agree with the density obtained by \citet{1981ApJ...245..154K}  and \citet{1992ApJ...390..536E}.

\subsubsection{Diagnostic diagrams \label{diagnostic}}
\begin{figure}
\centering
 \includegraphics[width=9cm]{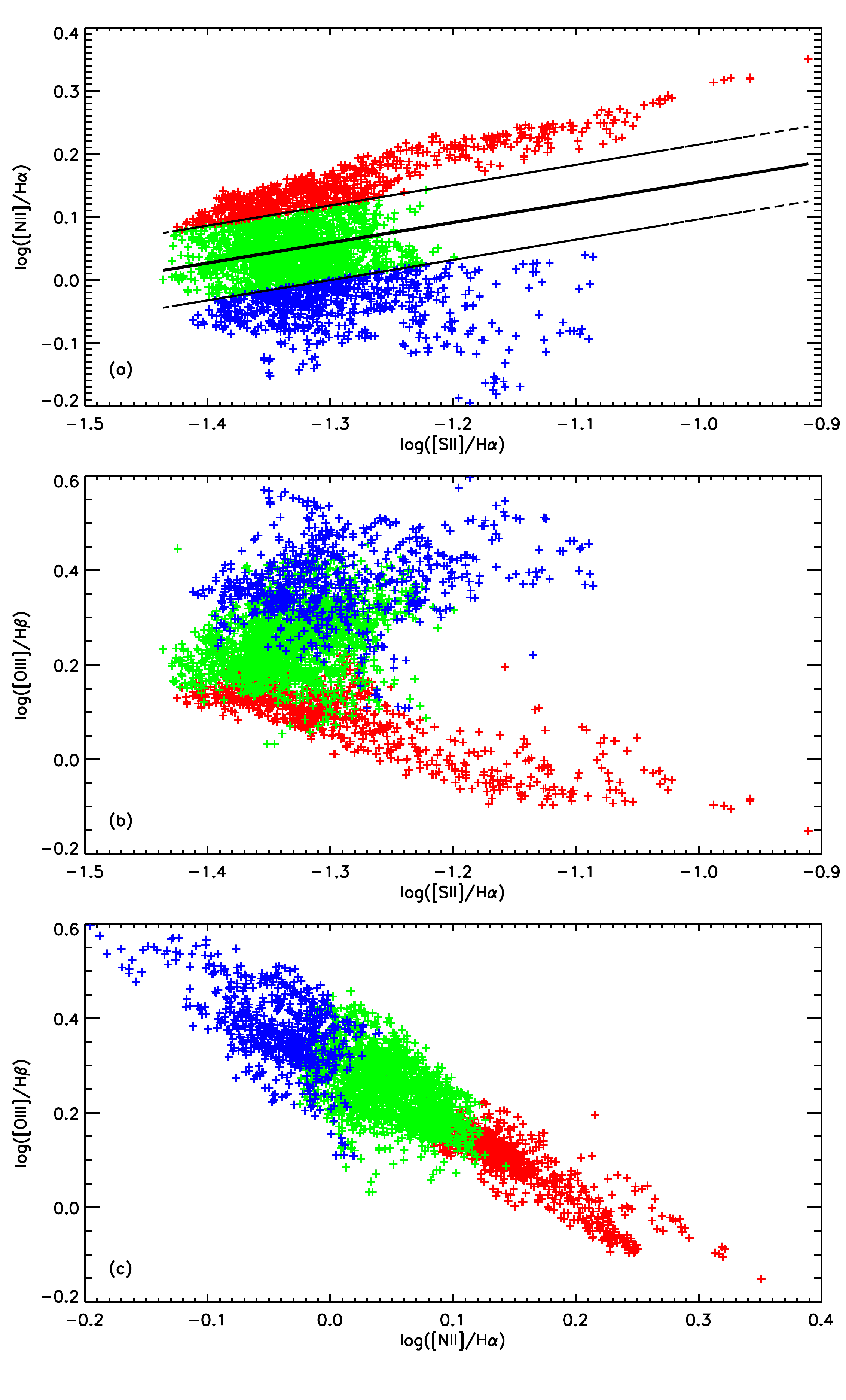}
\caption{Diagnostic diagrams with data from the 2D study of the X$-$ray pointing. From the top to the bottom: (a) [NII]$\lambda\lambda$6584/H${\alpha}$ vs [SII]$\lambda\lambda$6731/H${\alpha}$, (b) [OIII]$\lambda\lambda$5007/H${\beta}$ vs [SII]$\lambda\lambda$6731/H${\alpha}$, and (c) [OIII]$\lambda\lambda$5007/H${\beta}$ vs [NII]$\lambda\lambda$6584/H${\alpha}$. All the spaxels of the intensity maps are represented with crosses. Black lines in the first diagram are the fit performed and the $\pm$3$\sigma$ limits. Colours help us to locate points spatially in Fig. \ref{fig:zonas}: red corresponds to \emph{Zone A}, blue is \emph{Zone B}, and green represents \emph{Zone C} (see text for details).}
\label{fig:diagnostic2D}
\end{figure}

The excitation mechanisms and the ionization structure have been studied using diagnostic diagrams based on the ratio of strong emission lines: the classic BPT analysis \citep{1981PASP...93....5B}. Using the data from the 2D study of the X$-$ray zone, we generated maps for the necessary line ratios (with $S/N > 5$) and plotted all the spaxels in the three diagrams displayed in Fig. \ref{fig:diagnostic2D}.

\begin{figure}
\centering
 \includegraphics[width=6cm]{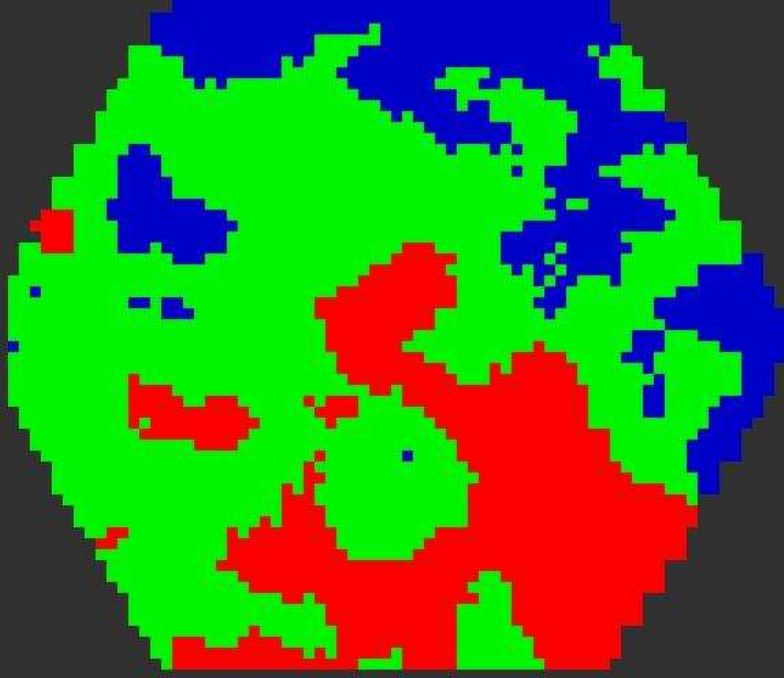}
\caption{PPAK FoV of the X$-$ray pointing with the different ionization zones defined by taking the tendencies in the BPT diagram of Fig. \ref{fig:diagnostic2D}\,a into account. Red corresponds to \emph{Zone A}, blue is \emph{Zone B}, and green represents \emph{Zone C}. Orientation as in Fig. \ref{fig:morphology_all}.}
\label{fig:zonas}
\end{figure}

The first diagram (Fig.\ref{fig:diagnostic2D}\,a) shows [NII]$\lambda\lambda$6584/H${\alpha}$ vs. [SII]$\lambda\lambda$6731/H${\alpha}$ (NvsS) and presents a bivaluate behaviour. We performed a reference fit to the points, paying attention to which spaxels are inside/outside the nominal limits of $\pm3\sigma$ from this fit. It can be seen that the relation between these two line ratios shows a significant scatter, especially for the points below the -3$\sigma$ line. We notice that spaxels above the +3$\sigma$ line show a linear$-$like correlation as could be expected from the ionization structure given the ionization degree of these ions. To understand these behaviours, we located the points of the diagram in the map differentiating between points inside, above and below the 3$\sigma$ limits of the global fit, to identify their spatial locations (see Fig.\ref{fig:zonas}). We found that points with a similar tendency in the BPT appear grouped in the PPAK map. The second diagram (Fig. \ref{fig:diagnostic2D}\,b) presents [OIII]$\lambda\lambda$5007/H${\beta}$ vs [SII]$\lambda\lambda$6731/H${\alpha}$ (OvsS), it shows a larger scatter with two tendencies. To understand how the zones defined above in Fig.\ref{fig:zonas} behave in this diagram, we identify pixels belonging to these zones with the same colours as in Fig. \ref{fig:diagnostic2D}\,a. Points with large scatter in NvsS (blue crosses) show scatter also in this diagnostic diagram, and points with a correlation in NvsS (red crosses) appear here anti$-$correlated. In the last diagram (Fig. \ref{fig:diagnostic2D}\,c) we can see the relation between [OIII]$\lambda\lambda$5007/H${\beta}$ and [NII]$\lambda\lambda$6584/H${\alpha}$ (OvsN), where we find a strong anti$-$correlation. Once again, we study the behaviour of the three regions defined in Fig.\ref{fig:zonas} in this diagram with the same colour code. In this case, the points appear anti$-$correlated for all the zones.

Taking the analysis and results explained above into account, we have defined three spatial zones in this pointing: \emph{Zone A} to the northwestern part of this pointing, corresponding to all the pixels above +3$\sigma$ in NvsS. We notice that in all the diagrams these spaxels show a normal and expected tendency. In addition, in this area the maximum emission in low excitation species is found. \emph{Zone B} is located to the southwest and corresponds to points below -3$\sigma$ in NvsS. In this case, the relations between the line ratios in the two first diagrams were rather anomalous and had a large dispersion. In the map this zone corresponds to the peak of the [OIII]$\lambda$5007\AA{} emission. Finally, we defined \emph{Zone C} by the points inside $\pm$3$\sigma$ limits of the fit in the diagram NvsS, which are located over the rest of the field and present a behaviour that is a mixture between the two other zones.\\

\subsubsection{Kinematics\label{kine}}
\begin{figure}
	  \centering
 \includegraphics[width=7cm]{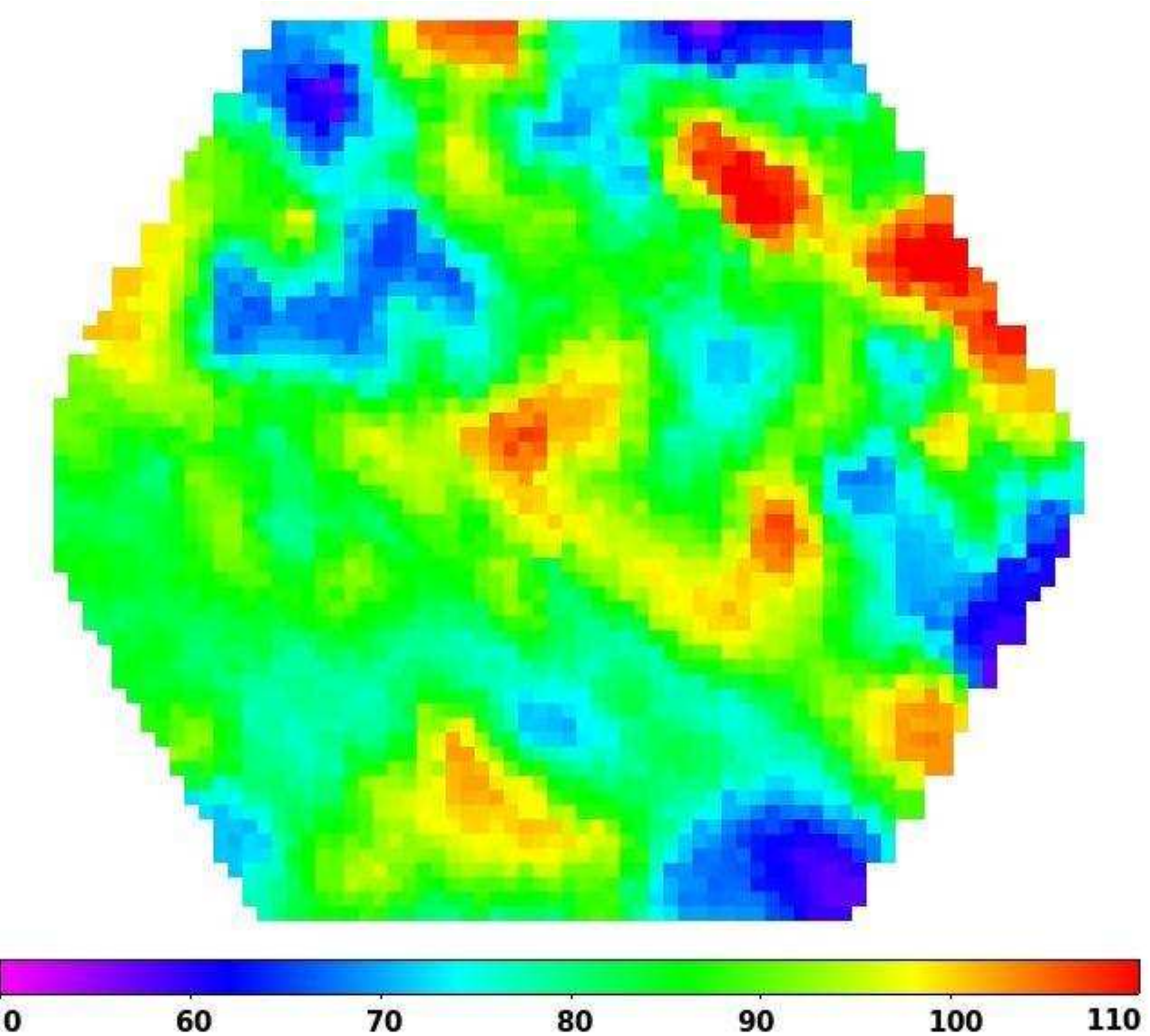}
		\caption{Radial velocity field derived for H${\alpha}$. Velocity ranges from 60 to 110 km~s$^{-1}$  with a mean value $\sim$84 km~s$^{-1}$. Orientation as in Fig. \ref{fig:morphology_all}.}
        \label{fig:ha_velocity}
  		\end{figure}

The radial velocity of the gas was derived using the central wavelength of the Gaussian fit in every pixel.  All the measured radial velocities were translated into the local standard of rest (LSR) and corrected for the Earth's motions using IRAF. In addition, a zero$-$velocity map was created using a \textquotedblleft sky\textquotedblright\, line, and this map gives us realistic information about the error in the wavelength calibration and the accuracy in the velocity that can be achieved. We concluded that differences in radial velocity below 8 km~s$^{-1}$ cannot be covered with our data. Figure \ref{fig:ha_velocity} shows the velocity field derived from the H${\alpha}$ emission line. The map has a mean value $\sim$84 km~s$^{-1}$, but it is not spatially homogeneous, because the southwest zone moves faster, reaching 110 km~s$^{-1}$,  while in the northwest zone the velocity is lower than 60 km~s$^{-1}$

\begin{figure}
	  \centering
	 \includegraphics[width=9cm]{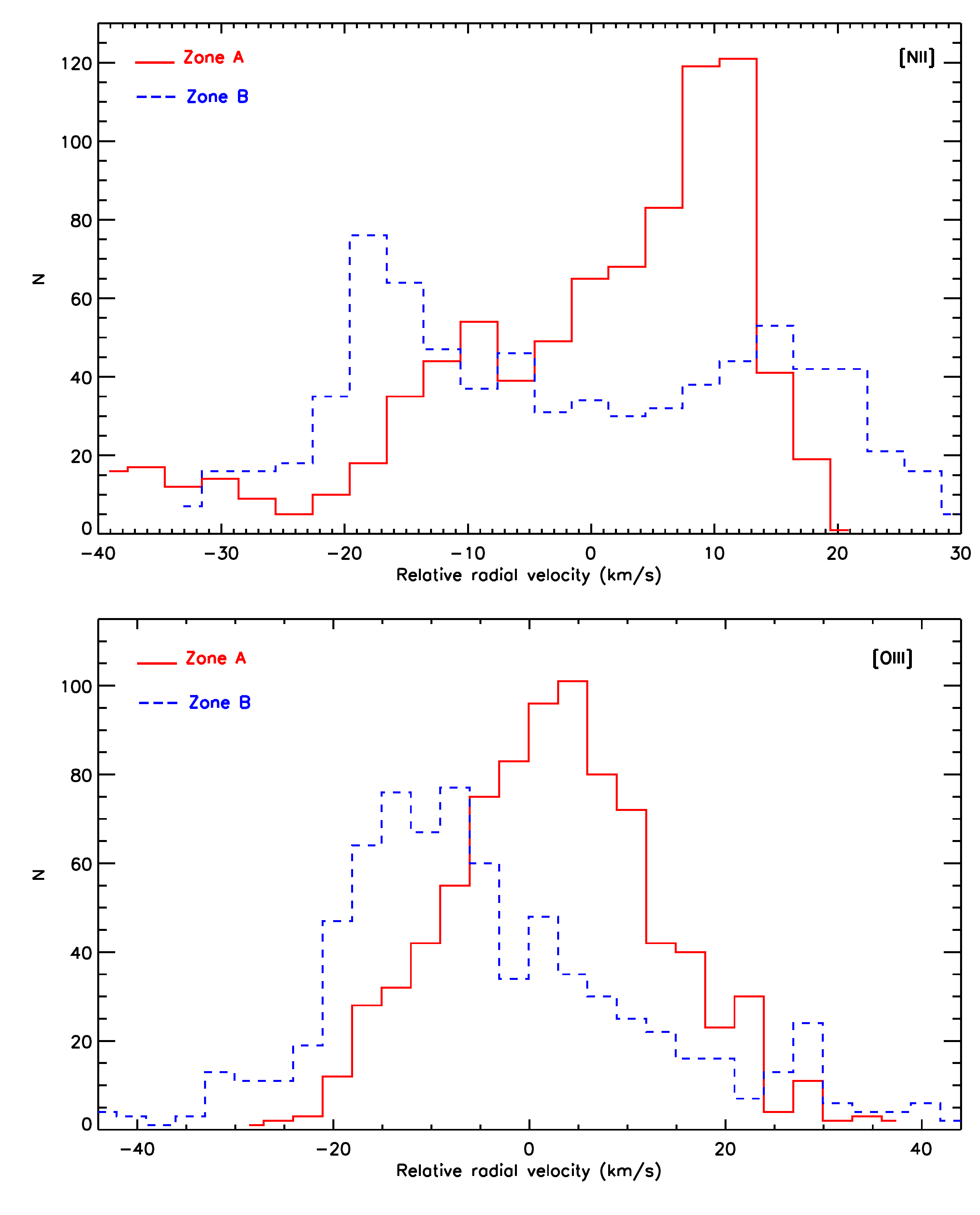}
		\caption{Statistical frequency distributions of the radial velocity of [NII]$\lambda$6584\AA{} and [OIII]$\lambda$5007\AA{}. Solid red lines indicate pixels from \emph{Zone A} and dashed blue lines from \emph{Zone B}. Velocities are given in units of km~s$^{-1}$ taking as zero the mean velocity measured for the emission line.}
        \label{fig:histograms}
  		\end{figure}

Taking advantage of the large number of pixels in our maps, we have produced statistical frequency distributions of the radial velocity. When we represent velocities from all the pixels, the histograms present a Gaussian distribution as expected. But, if pixels from \emph{Zones A} and \emph{B} are represented separately, a bimodal distribution is found in some relevant emission$-$lines like [NII]$\lambda$6584\AA{}. Figure \ref{fig:histograms} shows two histograms generated with a binning of 3 km~s$^{-1}$ for two emission lines differentiating the two defined zones. The [NII]$\lambda$6584\AA{} line presents two patterns: while \emph{Zone A} shows a distribution with a peak redshifted by 10 km~s$^{-1}$, \emph{Zone B} shows a bimodal distribution indicating a shell$-$like behaviour with pixels showing gas that is moving away and other pixels coming towards us.  On the other hand, in the second histogram, the [OIII]$\lambda$5007\AA{} line presents a single peak distribution in both zones and the velocity difference between the two peaks is $\sim$15 km~s$^{-1}$.\\

With this kinematic statistical study we support the idea that NGC\,6888 is a non$-$uniform nebula with shells and filaments moving with different velocities. Errors in the wavelength calibration and the instrument resolution of the observations prevent us from carrying out a more exhaustive analysis of the kinematics of the nebula in this work.

%________________________________________________________________

\section{ One$-$dimensional study \label{1dim}}		
\begin{figure}
		  \centering
	 \includegraphics[width=9cm]{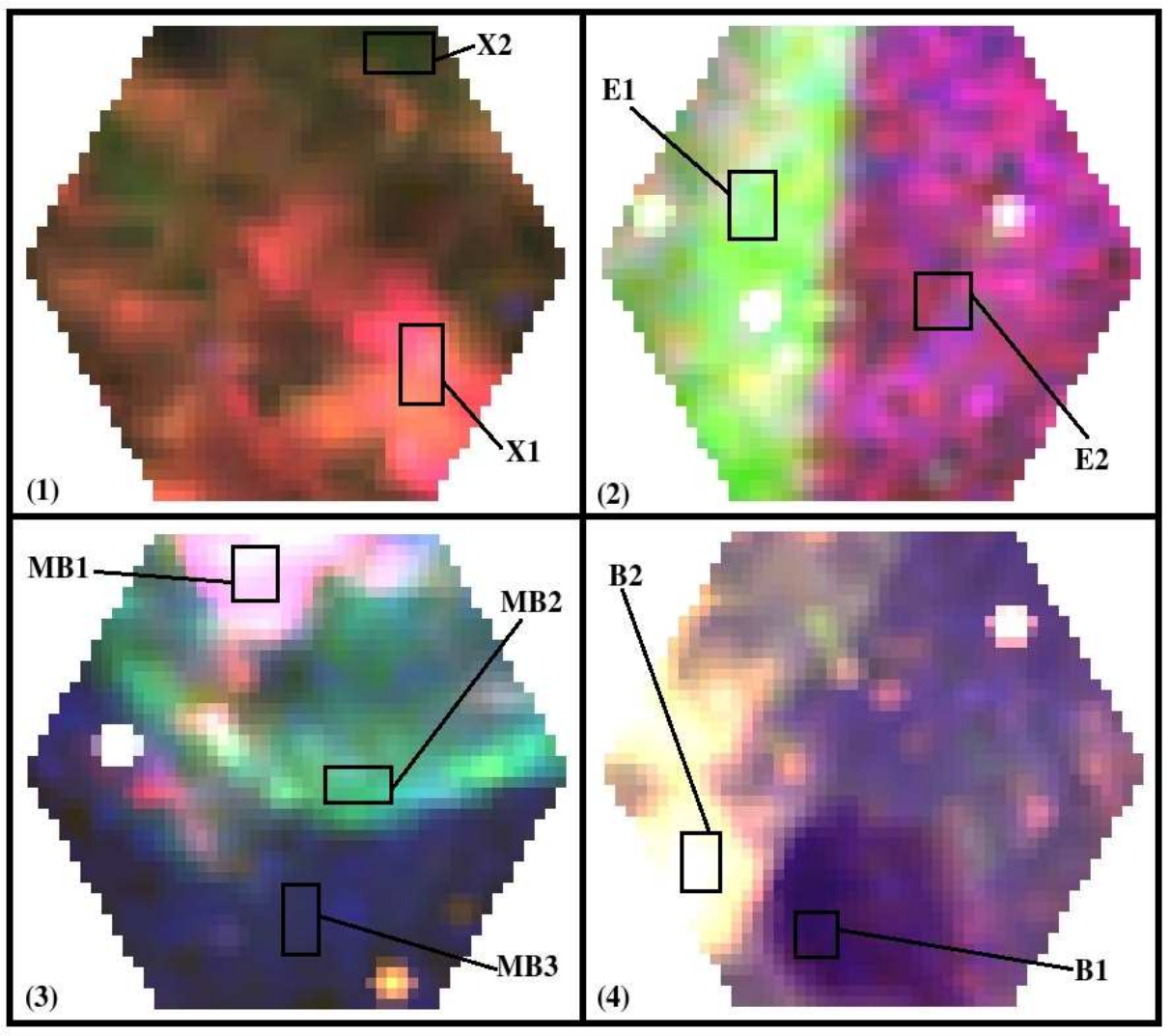}
		\caption{Displays using E3D of the four areas observed with PPAK. Boxes show regions where the integrated spectra were created: (1) X$-$ray zone, (2) Edge, (3) Mini$-$bubble, and (4) Bullet.  All the images are RGB$-$combinations of three wavelengths: red corresponds to the H${\alpha}$+[NII], green to [OIII], and blue to [OII]. Arbitrary scales. Orientation as in Fig.  \ref{fig:morphology_all}.}
     \label{fig:integratedspectra}
		\end{figure}

A different analysis of the IFS data can be carried out by combining the observed spectra in several fibres to produce integrated spectra in different regions of NGC\,6888. With this method we can use data without dithering exposures to obtain information over a broader spectral range and derive some physical properties that could not be obtained using a 2D approach. To decide which fibres are to be combined, we examined the appearance of each zone in emission lines at several wavelengths using E3D as shown in Fig. \ref{fig:integratedspectra} (see also Fig. \ref{fig:morphology_all}):\\
$-$ \textit{X$-$ray zone}: two integrated spectra were created in this pointing. The first spectrum (X1) corresponds to spaxels where the [NII]$\lambda$5755\AA{} line is visible. We have chosen this criterion because we were looking for emission lines useful for deriving the electron temperature. In addition, spaxels where [NII]$\lambda$5755\AA{} is bright enough coincide with the brightest low$-$excitation lines. To define the second region (X2), we selected spaxels placed in the area where we had a peak in the high ionization line emission ([OIII]). Moreover, X1 and X2 are in \emph{Zones A} and \emph{B}, respectively. \\
$-$ \textit{Edge zone}: we were interested in checking the different conditions and parameters on both sides of the nebular discontinuity observed in [OIII]$\lambda$5007\AA{}, so the first integrated spectrum was performed in the brighter zone to the left (E1) and the second one in the fainter part (E2). \\
$-$ \textit{Mini$-$bubble}: we created integrated spectra over three of the features seen. Firstly, we chose fibres emitting in [NII]$\lambda$5755\AA{} (MB1). Secondly, a bright zone in [OIII]$\lambda$5007\AA{} but without [NII]$\lambda$5755\AA{} emission (MB2). Finally, we defined the zone with low level of emission at all the wavelengths (MB3). \\
$-$ \textit{Bullet zone}: to check the nature of the \emph{bullet} candidate, we created one integrated spectrum over the dark zone (B1). In addition, as a representative example of the emission of the inner area of the nebula, we extracted a spectrum (B2) in this brighter region but outside of the bullet. \\

\begin{figure}
		  \centering		
		 \includegraphics[width=9cm]{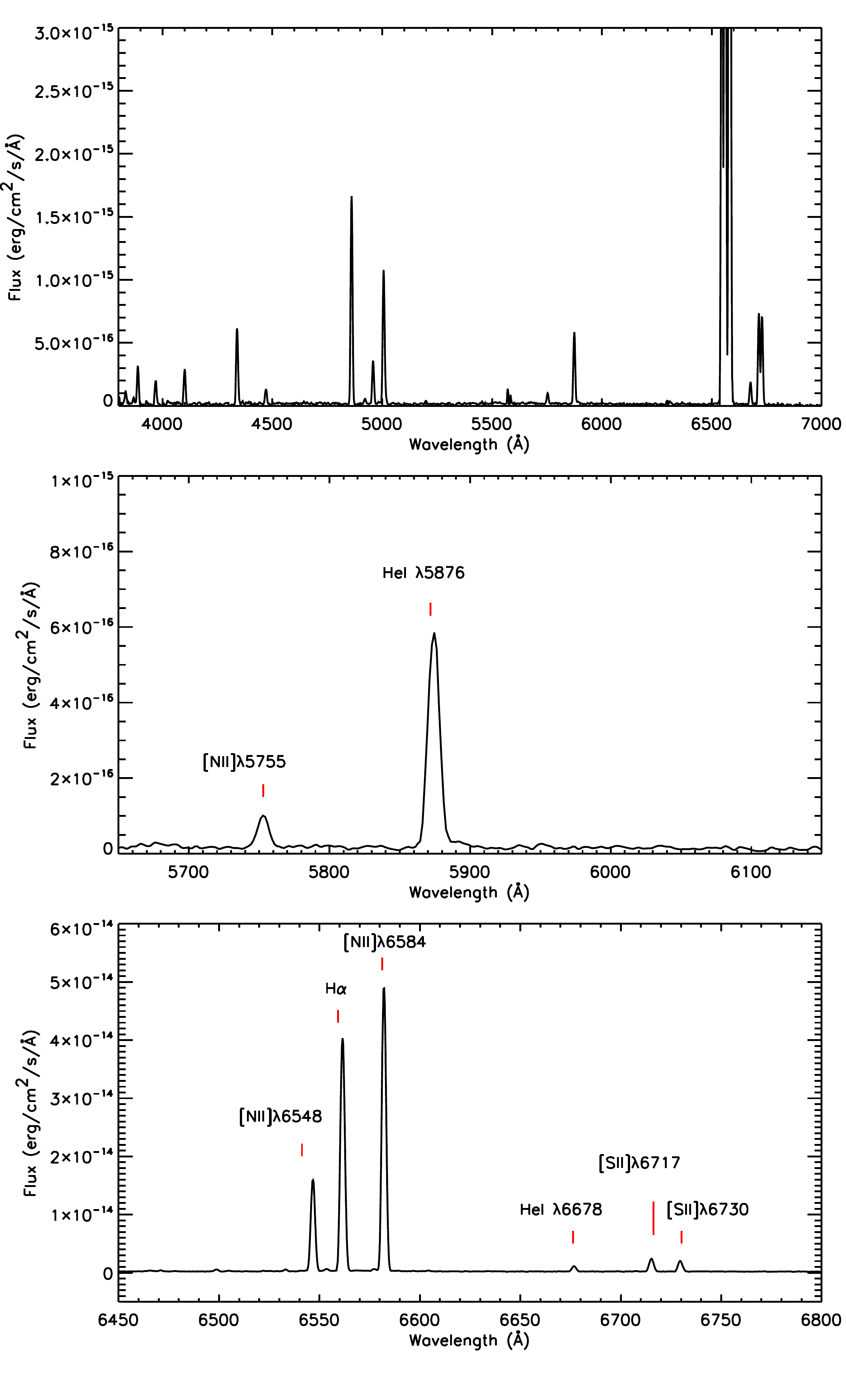}
		\caption{Integrated spectra corresponding to the low ionization area of the X$-$ray zone (X1). Top: the whole spectrum in the V300 grating. Middle: a zoom centred on the [NII]$\lambda$5755\AA{} and HeI$\lambda$5876\AA{} emission lines. Bottom: the R1200 red set up.}
      \label{fig:integ}
 		\end{figure}

Using these definitions, several sets of fibres were selected and combined for each grating in all the zones to obtain their integrated spectra, which were analysed to derive their spectroscopic properties in the whole spectral range of the observations. For observations where a dithering scheme was adopted we combined the fibres of all the exposures. At the end, we obtained 19 1D spectra considering the different gratings. Figure \ref{fig:integ} shows an example of integrated spectra in zone X1.

        \subsection{Integrated properties\label{int_prop}}
The emission$-$line fluxes were measured using the SPLOT task in IRAF. Lines show a degree of asymmetry in their profiles, so we did not perform Gaussian fits The fluxes were derived measuring areas below the lines, however, partially blended lines, such as H${\alpha}$ and [NII]$\lambda\lambda$6584,6584\AA{} in the V300 grating, were deblended with two or three Gaussian profiles to measure the line fluxes separately. The statistical errors associated with the observed fluxes were calculated using the expression
\begin{equation}
	\sigma_{\mathrm{1}}=\sigma_{\mathrm{c}} N^{1/2} [1+EW/(N\Delta)]^{1/2}
\end{equation}
where $\sigma_{\mathrm{1}}$ represents the error in the observed line flux, N is the number of pixels used to measure the line, EW the line equivalent width, $\sigma_{\mathrm{c}} $ the standard deviation of the continuum in a box near the line, and $\Delta$ represents the dispersion in \AA{}/pix \citep{2003MNRAS.346..105P}.\\

Table \ref{table:all_lines} lists the emission lines measured in every zone labelled with its standard identification. The estimated fluxes and errors were normalized to $F(H{\beta})=100$. Line intensities were reddening$-$corrected using the same procedure as described in Sect. \ref{extinc}. In this case, other HI Balmer lines with $S/N>5$ were measured in the integrated spectra and therefore used to derive the reddening coefficient. In all the spectra, H${\alpha}$,  H${\beta}$, H${\gamma}$,  and H${\delta}$ were measured with enough S/N and used to estimate c(H${\beta}$), with the exception of the pointing B1 in which H${\gamma}$  and H${\delta}$ were very faint. The reddening coefficients derived in zones X1 and X2 agree with the values obtained in the 2D$-$study in the X$-$ray pointing using only H${\alpha}$ and H${\beta}$. %Figure \ref{fig:fitC} shows one of the the fits performed to obtain the reddening coefficient using Balmer lines.
The values obtained for c(H${\beta}$) are presented in the last row of Table \ref{table:all_lines}. Errors in the emission$-$line intensities were derived by propagating the observational errors in the fluxes and the reddening constant uncertainties.

In regions X1, X2, E1, and E2, some lines were measured more than once since they appear in exposures with different gratings. In these cases, we calculated the corresponding intensity as the mean weighted by the error in each zone.\\

To obtain the physical conditions of the gas, we performed an iterative process until achieving agreement between electron temperature and electron density. Table \ref{table:paramyabun} shows the values for T$_{\mathrm{e}}$ and n$_{\mathrm{e}}$ derived for each zone.

With the [SII]$\lambda\lambda$6717/6731 line ratio, electron densities (n$_{\mathrm{e}}$) were obtained using the package TEMDEN of IRAF. We derived differences in the electron density of up to 350 cm$^{-3}$ between the studied zones. The results from the X-ray zones are coherent with the values obtained in the bi-dimensional analysis where densities were higher in the northwest of our field of view that corresponds to the integrated spectrum X1.  On the other hand, values obtained on the edge of the nebula (E1 and E2) and in the less bright zones of the mini$-$bubble (MB2 and MB3) are lower than the densities derived in other regions (always $n_{\mathrm{e}}<100$ cm$^{-3}$).

Electron temperature, T$_{\mathrm{e}}$, can be derived using the line ratios

\begin{equation}
R_{\mathrm{O3}}={I([\mathrm{OIII}]\lambda 4959)+I([\mathrm{OIII}]\lambda 5007) \over I([\mathrm{OIII}] \lambda 4363)}
\end{equation}
and

\begin{equation}
R_{\mathrm{N2}}={I([\mathrm{NII}]\lambda 6548)+I([\mathrm{NII}]\lambda 6584) \over I([\mathrm{NII}]\lambda 5755)}\,.
\end{equation}

The [OIII]$\lambda$4363\AA{} line was blended with the \textquotedblleft sky\textquotedblright\, line HgI$\lambda$4358\AA{}, and therefore we cannot compute T$_{\mathrm{e}}$([OIII]) directly. In contrast, the [NII]$\lambda$5755\AA{} auroral line was measured with enough confidence in zones X1, MB1, and B2, and direct estimates of T$_{\mathrm{e}}$([NII]) were obtained from the R$_{\mathrm{N2}}$ ratio. In the rest of zones (X2, E1, E2, MB2, MB3, and B1), for which this line was not bright enough to be measured with confidence, we resorted to the empirical flux$-$flux relations from \citet{2005A&A...436L...1P} that relate the faint auroral lines with other strong nebular lines. To find out what was the best relation to our particular case, we used several fits proposed by Pilyugin over the zone X1, for which R$_{\mathrm{N2}}$ is well measured, and selected which relation represented our results better. We chose the relation $Q_{\mathrm{NII}}=f(R_{\mathrm{2}},P)$ from \citet{2007MNRAS.375..685P} to estimate the electron temperature T$_{\mathrm{e}}$([NII]) and checked that this relation gives consistent results with our measurements of the line in the X1 zone. Later, from T$_{\mathrm{e}}$([NII]), we derived T$_{\mathrm{e}}$([OIII]) and T$_{\mathrm{e}}$([OII]) in each zone by using the relations based on models proposed by \citet{2009MNRAS.398..949P} and \citet{2003MNRAS.346..105P}, respectively. Derived temperatures range from $\sim$7700~K to $\sim$10\,200~K over the nebula, which is consistent with the previous studies of \citet{1981ApJ...245..154K} and \citet{1992ApJ...390..536E}.

        \subsection{Chemical abundances\label{abund}}
Considering the electron density and temperatures estimated for each ionization zone, we obtained ionic abundances from the forbidden$-$to$-$hydrogen emission$-$line ratios using the functional forms given by \citet{2008MNRAS.383..209H}, which are based on the package IONIC of IRAF. To determine the singly ionized helium abundance, we used the equations proposed by \citet{2004ApJ...617...29O}. We defined two temperatures: T$_{\mathrm{e}}$([NII]) as temperature representative for the low ionization ions, S$^{+}$, N$^{+}$, and O$^{+}$, and T$_{e}$([OIII]) for the high ionization species, O$^{2+}$ and Ne$^{2+}$, and for deriving the abundance of He$^{+}$. All the ionic abundances are presented in Table \ref{table:paramyabun} with their corresponding errors.

Furthermore, in Table \ref{table:paramyabun} we show the total abundances derived as follows. The O/H has been obtained by directly adding the two ionic abundances (O$^{+}$ and O$^{2+}$). The same procedure was used for He/H but only with He$^{+}$ because He$^{2+}$ is not seen in our observations, the value chosen of HeI was a mean weighted by the errors of the three HeI observed lines (HeI$\lambda$4471\AA{}, HeI$\lambda$5876\AA{}, and HeI$\lambda$6678\AA{}). The N/O can be approximated to N$^{+}$/O$^{+}$. Finally, to derive the total Ne/H abundance, ionization correction factors (ICF) are required to take the unseen ionization stages into account, we used the ICF expression for Ne from \citet{2007MNRAS.381..125P}. Taking \citet{2007ApJ...662...15I} as reference, we estimated $ICF(He)\sim1$ for all the zones, so correction for the unseen neutral helium was not necessary.

Mainly, three results in the total abundances are interesting. First, we see that nearly all the zones present an oxygen abundance slightly below the solar value, taking the solar abundance from \citet{2009ARA&A..47..481A} as reference . This effect was reported above for the areas observed by \citet{1981ApJ...245..154K} and \citet{1992ApJ...390..536E}. On the other hand, these authors derived an overabundance of nitrogen. For some of our zones (E1, E2, MB2, MB3, and B1), we derived a nitrogen abundance similar to the solar value, however in other zones the N/H appears enhanced by a factor of 6, or even 8 in the X1 zone. Helium presents an enrichment in most of the integrated zones too. (MB3, E2, and B1 do not show the HeI line.) One of the most striking results is the important differences in the N/O between the different zones, mainly due to the large N/H variation and not to differences in O/H. This could be understood by assuming that we are seeing structures of the nebula formed in different stages of the evolution of the WR star, each one with its own ionization degree, physical properties, and chemical patterns. This idea is analysed in Sect. \ref{discussion}. Furthermore, we calculated the expected values in the ISM at the galactocentric radius of  NGC\,6888, considering the galactic chemical abundance radial gradients. Taking \citet{2011ApJ...738...27B} as reference, we derived $12+\log \mathrm{(O/H)}=8.6\pm0.05$ and $\log \mathrm{(N/O)}=-1.26\pm0.26$ from \citet{2005ApJ...623..213C}. Our inferred abundances for E2, MB3, and B1 are consistent with these expected values (within the errors).

        \subsection{Photoionization models\label{grids}}
To investigate and consolidate both ionization parameter (U) and chemical abundances in the different pointings of the nebula, we performed a grid of photoionization models made with the code Cloudy v.8.0 \citep{1998PASP..110..761F}. These models are based on the grid described by \citet{2009MNRAS.398..949P}, using one single stellar atmosphere of WR, in particular, a WN star model \citep{2002MNRAS.337.1309S}. Our grids assume an effective temperature of the star of 50~000 K \citep{1990ASPC....7..135R,1993A&A...272..299E}, with a metallicity of $Z = 0.008$. Taking the solar abundances reported by \citet{2009ARA&A..47..481A} ($12+\log \mathrm{(O/H)} = 8.69$, $Z = 0.013$) as a reference, the assumed metallicity implies an oxygen abundance $12+\log \mathrm{(O/H)})=8.47$, which is the closest value to the O/H abundances derived in this work for NGC\,6888. These models vary the ionization parameter and the nitrogen$-$to$-$oxygen ratio as free parameters, which are very relevant for our purpose. We considered values $\log(U) = -2.5, -3.0$, and $-3.5$, and $\log \mathrm{(N/O)}$ from -1.5 to 0.0 (in steps of 0.25).

\begin{figure}
		  \centering
		 \includegraphics[width=9cm]{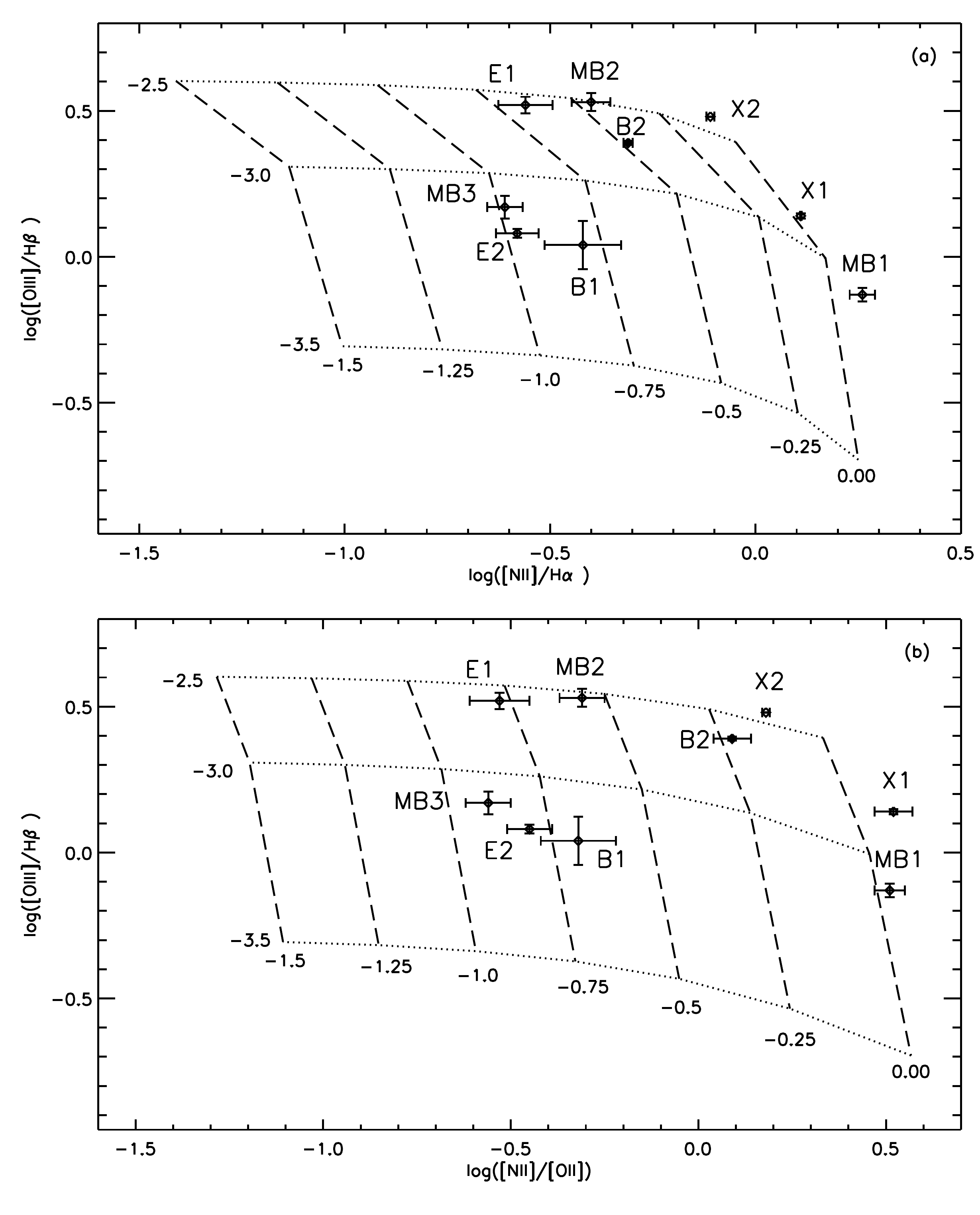}
		\caption{Diagnostic diagrams containing the results of our photoionization models and observed data. (a) [OIII]$\lambda\lambda$5007/H${\beta}$ vs [NII]$\lambda\lambda$6584/H${\alpha}$ and (b) [OIII]$\lambda\lambda$5007/H${\beta}$ vs [NII]$\lambda$6584/[OII]$\lambda\lambda$3726,3729. In all the diagrams, the dotted lines connect points with the same ionization parameter. The dashed lines connect curves of equal $\log \mathrm{(N/O)}$.  The respective values are indicated at the beginning of curves. For all the models we assumed $12+\log \mathrm{(O/H)})=8.47$. Points are values of integrated spectra placed according to their intensities ratios (named as in Table \ref{table:all_lines} and Fig \ref{fig:integratedspectra}).}
      \label{fig:grids_and_data}
 		\end{figure}

We compared selected emission-line ratios of the integrated spectra with models in two diagnostic diagrams: [OIII]$\lambda\lambda$5007/H${\beta}$ vs [NII]$\lambda\lambda$6584/H${\alpha}$ and [OIII]$\lambda\lambda$5007/H${\beta}$ vs [NII]$\lambda$6584/[OII]$\lambda\lambda$3726,3729 \citep{2000ApJ...542..224D}. As can be seen in Fig. \ref{fig:grids_and_data}, for each zone, the values of $\log \mathrm{(N/O)}$ obtained with the comparison always agree with abundances calculated with our measures. For each integrated spectra, the ionization parameter is similar in the two diagrams: zones X2, E1, and MB2 have the greater U; zones X1 and B2 have intermediate values; and the paremeters are the lowest in zones E2, MB1, MB3, and B1. \footnote{Models with an effective temperature of 45~000 K and WM-Basic stellar  atmosphere \citep{2001A&A...375..161P} also gave satisfactory results.}

%________________________________________________________________

\section{Discussion\label{discussion}}
\begin{figure}
		  \centering
		 \includegraphics[width=7cm]{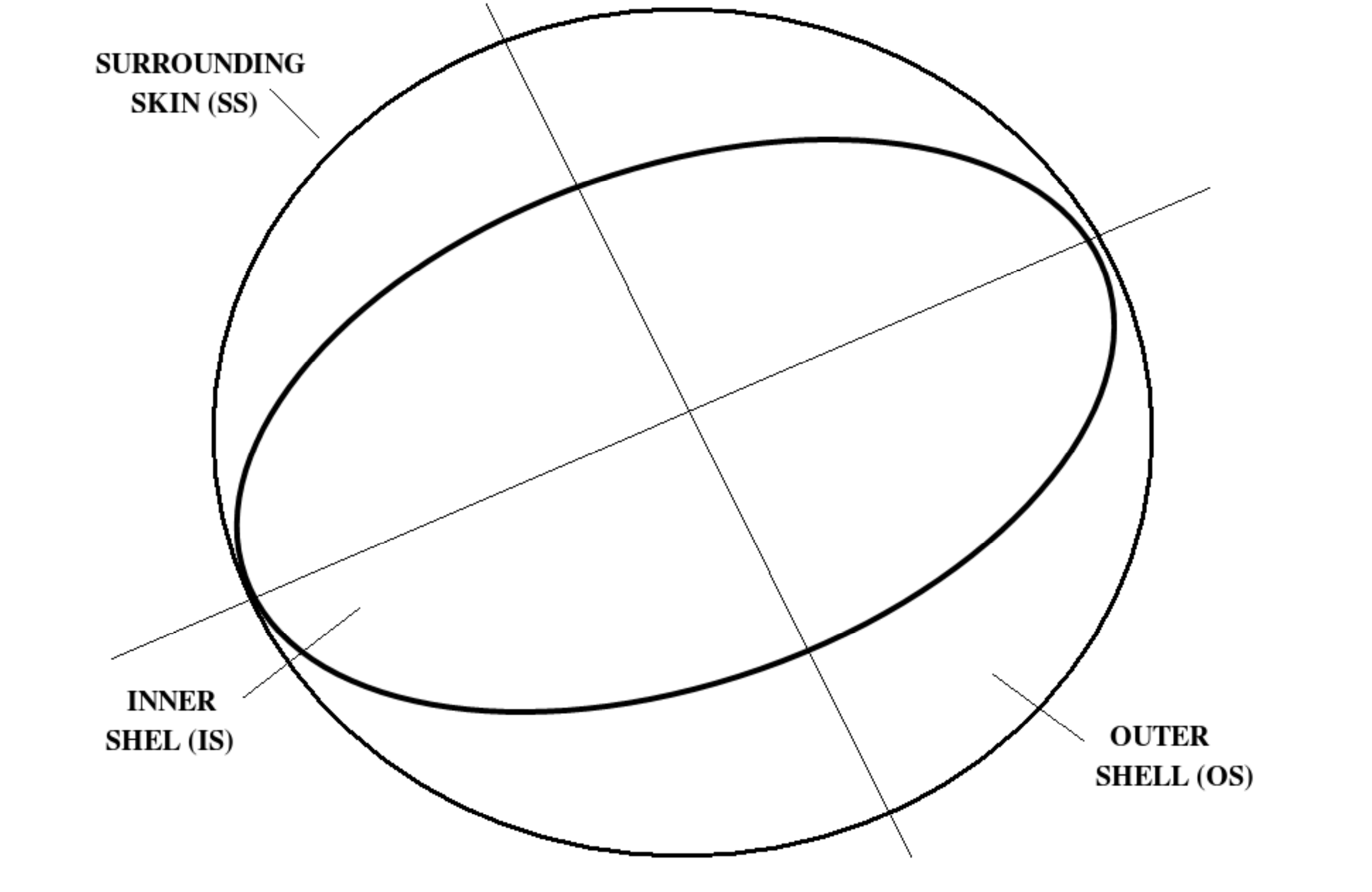}
\caption{Sketch showing the structure of NGC 6888 with two main shells: one inner (IS) and the other outer (OS).}
      \label{fig:scheme}
 		\end{figure}

To develop a scheme of the internal structure of NGC\,6888, we put together all the information derived from the 2D maps and the nine integrated spectra. Furthermore, we used the narrow$-$band images from the INT$-$WFC shown in Sect. \ref{intro} to have a global overview of the nebula and to locate regions of different observed conditions. The proposed scheme consists of a shell structure (close to \textquotedblleft onion$-$like\textquotedblright) but allowing irregular and/or broken shells. We found, at least, two shells with different physical, kinematical, and chemical properties: an outer shell (OS) and an inner shell (IS). In addition, a more external and fainter structure is seen surrounding the whole nebula like a thin \emph{skin}; we call it \textit{surrounding skin} (SS). A basic sketch of this scheme is displayed in Fig \ref{fig:scheme}.

Our interpretation of the structure of NGC\,6888 is supported by comparison to models of WR ring evolution. In particular, there is very good agreement with the simulation by \citet{1996A&A...316..133G} (hereinafter GS96) for the interaction of a 35 $M_{\sun}$ star with the circumstellar medium over the entire stellar lifetime in the case of the \emph{slow} RSG wind ($\sim$15 km~s$^{-1}$). Their simulations start at the MS stage, when the stellar winds sweep up the ISM, forming a hot bubble surrounded by a thin shell. Later, in the red supergiant (RSG) phase, the star is characterized by a slow and very dense wind that shocks with the MS bubble, creating a shell (RSG shell). Finally, a faster wind from the WR phase sweeps up the RSG wind material into a new shell; however, the propagation of this WR shell is faster, and it soon arrives at the RSG shell. After the collision, clumps and dense material of the WR shell are propelled by their high inertia, crossing the RSG shell as bullets forming blow$-$outs. The final stage in the GS96 simulations (Fig. 7 in that paper) shows that the bubble formed in the main sequence is, eventually, broken out as consequence of the RSG$-$shell shocked by the WR shell and its expansion.\\

   			    \subsection{The inner shell (IS)\label{IS}}
In the proposed scheme, the central part of NGC\,6888 is an ellipsoidal shell surrounding the WR star with a high surface brightness, as can be seen in the  INT image (Fig. \ref{fig:zone}).  This shape agrees with the scheme proposed previously by \citet{1988SvAL...14..385L}. Based on interferometric observations, they derived a spatial model in which NGC\,6888 is a prolate ellipsoid of revolution  ($a/b~\sim 0.5$) whose major axis is inclined 20$^{\circ}$ $-$ 40$^{\circ}$ to the projection plane.

In our data, the information about this shell was obtained from the integrated spectra X1, X2, MB1 and B2. From the analysis of the properties of these four zones, we propose that this shell is not  a complete shell, but a broken structure, with holes, filaments and bubbles. One evidence pointing to this are the variations of the electron density always $>$100 cm$^{-3}$ ranging from 106 to 360  cm$^{-3}$ along the defined IS.  Also, differences in the electron temperature and in the chemical composition show inhomogeneities between different zones of this shell.

One of the more important results for the IS is the strong overabundance in N/H by a factor between 5 and 9 with respect to the solar abundance and by a factor up to $\sim$11 with respect to other regions of the nebula. The enriched N/H  found varies from $12+\log \mathrm{(N/H)}=8.29$ to $8.72$ in the integrated spectra included in this shell, supporting the idea of a patchy and incomplete structure. On the other hand, the O/H abundance is found to be slightly deficient, suggesting that some oxygen could have been processed into N mainly via the ON cycle. In the case of helium, we obtained a value of up to He/H=0.18 in this shell, indicating  a strong overabundance of this element when we compare it with the solar value He/H=0.09 \citep{2009ARA&A..47..481A}. These facts give us clues to the stage in the star evolution in which this shell originated.

\begin{figure}
		  \centering
		 \includegraphics[width=9cm]{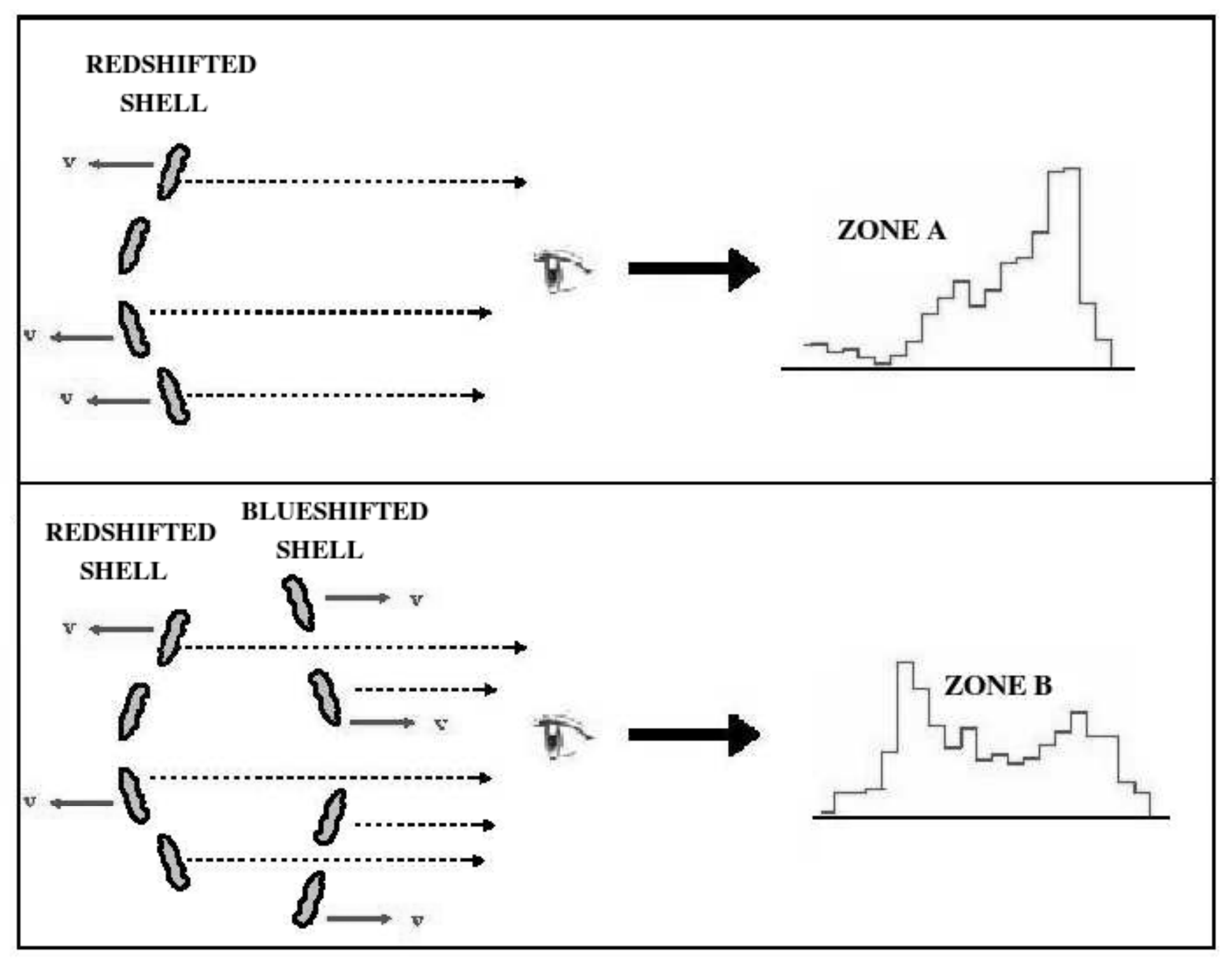}
\caption{Sketch showing the proposed broken inner shell. When we look at \emph{Zone A}, we only receive light from the furthest shell that is moving away from us (the histogram shows a redshifted peak); on the other hand, when we look at \emph{Zone B}, we see gas belonging to both shells (blueshifted or redshifted) but not simultaneously or necessarily in the same spaxels.}
      \label{fig:inner}
 		\end{figure}

We have additional information about the IS thanks to the observations with IFS because the X$-$ray pointing (with dithering and full datacubes) is over this region. In Sect \ref{diagnostic} we defined two zones, \emph{Zones A} and \emph{B}, associated with the emission of the low and high ionization ions, respectively, and showing very different trends in the diagnostic diagrams. Later, we found two behaviours in the radial velocities associated to these zones. In particular, the statistical frequency distributions of velocity (Fig. \ref{fig:histograms}) are consistent with a broken shell scheme: the [NII]$\lambda$6584\AA{} velocity histogram in \emph{Zone A} suggests only one redshifted shell going away from us, while in \emph{Zone B} the bimodal velocity distribution found can be understood if the shell is moving away in some places and in others is coming towards us. Rarely are the two behaviours observed simultaneously, because we do not see a central and dominant velocity component in the histograms. This kinematical study supports the idea of an intrinsically inhomogeneous shell in expansion, more obvious in the nitrogen than in the oxygen emission. Figure \ref{fig:inner} shows a sketch of how the scheme of broken$-$shell works and what we would see for each zone. \\

Although in our work we cannot consider the 2D study of the whole nebula, including all the spatial information, we propose that the most internal shell of NGC\,6888 should present a structure with inhomogeneities in the physical properties and abundances. Differences in the chemical composition could be understood by attending to the phase in the evolution of the central star when the shell was formed; in particular and in comparison with GS96 simulations, we argue that the IS material corresponds to the red supergiant (RSG) and WR shells shocked expanding in a low$-$density medium. This would explain its fragmented and filamentary structure and the high emission of H${\alpha}$. Also, the unusual abundances found  indicate that this shell is mostly composed of stellar material ejected from the central star when it became a WR star and later probably mixed with material from the RSG stage. This idea is also supported by the recent simulation performed by \citet{2011ApJ...737..100T}. They find that the excess of N/O in NGC\,6888 can be explained and reproduced by a stellar evolution model with rotation \citep{2003A&A...404..975M} for an initial stellar mass of 37 $M_{\mathrm{\sun}}$.

  \subsubsection{Can we detect the footprint of shocks?\label{shock}}
The previous observations of X ray emission detected in this region and the theoretical simulations (GS96) lead us to consider that on the north edge of the major axis of this IS, a collision between the WR and RSG shells happened during the star's lifetime. To check that there are spectral signatures of shocks, we resorted to the models presented by \citet{2008ApJS..178...20A}. Their code, MAPPINGS III, gives an extensive library of radiative shock models covering a wide range of abundances and velocities. First, we had the maps of the emission lines ratios created in our 2D study; then, we generated a grid of shock models with different parameters using libraries from \citet{2008ApJS..178...20A}; finally, we overplotted all the spaxels from the maps on the theoretical grids in the diagnostic diagrams to find out the consistency with models.

\begin{figure}
		  \centering
	 \includegraphics[width=9cm]{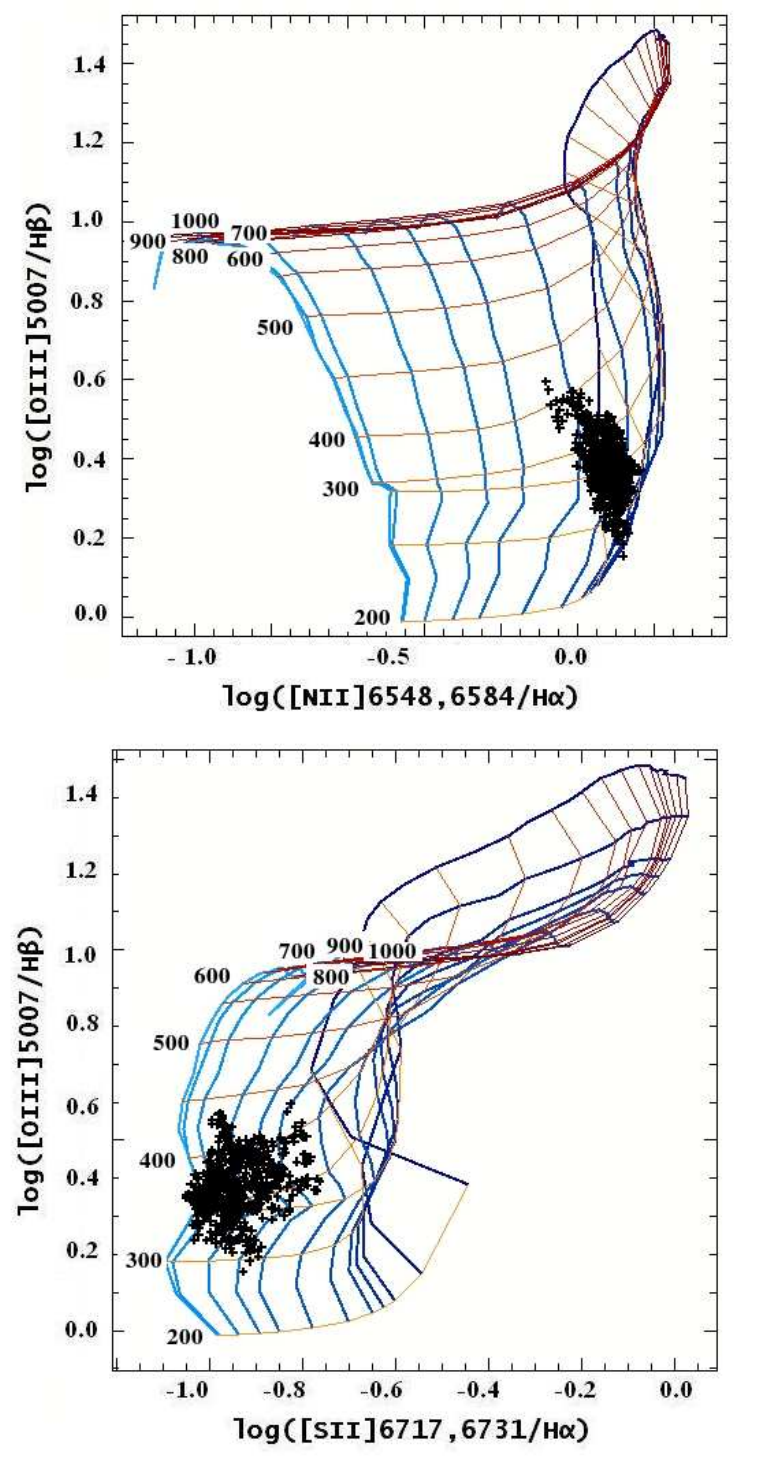}
 \caption{Diagnostic diagrams: [OIII]$\lambda\lambda$5007/H${\beta}$ vs [NII]$\lambda\lambda$6548,6584/H${\alpha}$ top, and [OIII]$\lambda\lambda$5007/H${\beta}$ vs [SII]$\lambda\lambda$6717,6731/H${\alpha}$ bottom. Grids represent the prediction from the shock models using libraries from \citet{2008ApJS..178...20A}. Numbers over the lines indicate the value of the shock velocity represented by red lines, while the blue lines are models with equal magnetic field. Crosses are the pixels from NGC\,6888 maps belonging to the \emph{Zone B}. Only data from \emph{Zone B} are represented well in both diagrams by a model with shock and precursor, velocities from 250 to 400 km~s$^{-1}$, $n=1000~cm^{-3}$, and twice solar abundance.}
      \label{fig:allen}
 		\end{figure}

In a first attempt, we plotted all the pixels of our maps in two diagnostic diagrams ([OIII]$\lambda\lambda$5007/H${\beta}$ vs [NII]$\lambda\lambda$6548,6584/H${\alpha}$ and [OIII]$\lambda \lambda$5007/H${\beta}$ vs [SII]$\lambda\lambda$6717,6731/H${\alpha}$), but we failed to find a model that represented all the observational data in both diagrams well. Later, we studied only the region with a striking behaviour in Figs. \ref{fig:diagnostic2D} and \ref{fig:zonas} (called \emph{Zone B}). When we plotted only the spaxels that belong to this region, we found that they can be fit with one of the grids of models. Figure \ref{fig:allen} represents the two diagnostic diagrams showing the pixels from the \emph{Zone B}, together with the grid of shock models.

We found that our data of \emph{Zone B} can be represented well by a model in which shock and precursor are involved assuming n=1000 cm$^{-3}$ and with twice solar abundances. In both diagrams we find that the models that fit better have shock velocities from 250 to 400 km~s$^{-1}$. The agreement of our data with the predictions of shock models with twice solar abundances can be understood by attending to the strong overabundance of nitrogen. Nevertheless, we should bear in mind that in our 1D study the line ratios of this region (integrated spectra X2) are also consistent with the tailored photoionization model presented in Sect. \ref{grids}. An alternative scenario can be considered that is similar to the ionization of \textquotedblleft pillars of creations\textquotedblright \,structures \citep{2011MNRAS.tmp.1984E}. In this scenario, altered line ratios could be produced as a consequence of the partial observation of ionized to neutral surfaces in the nebula.

As a conclusion, considering the shock models, the X$-$ray emission and the stellar evolution simulations, we cannot rule out the presence of a shock, so if this event has happened, the analysis of our data indicates that the most likely place to find it is \emph{Zone B} of the X$-$ray pointing. It would be interesting to check this with new observations of the other side of the major axis (southwest of NGC\,6888).

			    \subsection{The outer shell (OS)\label{OS}}
Studying the integrated spectra E1 and MB2, we found several differences in the physical properties in comparison with other zones that lead us to think that they  belong to a different shell. In our nebular scheme, we propose an intermediate shell  placed between the SS and the IS. The morphological study made in the bi$-$dimensional analysis (Sect. \ref{morphology})  supports the existence of  different shells or spatial inhomogeneities in the structure. In particular, the images in the second row of  Fig. \ref{fig:morphology_all}, corresponding to the Edge pointing, show a clear discontinuity in the spatial distribution of the emission of the [OIII]$\lambda$5007\AA{} line that is not visible in H${\alpha}$ nor [OII]$\lambda$3727\AA{}. The integrated spectrum E1 was created from the emission to the right of this discontinuity. Something similar happens in the mini$-$bubble pointing (third row in Fig. \ref{fig:morphology_all}) where a well$-$shaped semicircular structure (where the MB2 spectrum was extracted) is  seen only in the [OIII] emission line.

Taking the previous observations of NGC\,6888 into account \citep{2000AJ....120.2670G,2000AJ....119.2991M}, it is very likely that the OS has an apparent spherical shape, at least in the [OIII]$\lambda$5007\AA{} emission line.  The INT$-$WFC image (Fig. \ref{fig:zone}) agrees with this morphology: it can be seen that the oxygen distribution is not as elliptical as the other elements, showing some extended emission beyond the end of the minor axis  of the IS.

This bubble is surrounding the inner shell and looks dominant in double ionized oxygen. Intensities measured of the emission lines are weaker here than in the inner shell; nevertheless, the [OIII]$\lambda$5007\AA{} line has the maximum flux in comparison with the rest of the integrated spectra. Here we estimate the highest electron temperatures of the whole sampled nebula ($>$10\,000~K). The expansion of the shell could explain the electron density found here, always lower than 100 cm$^{-3}$. The chemical analysis shows that the He/H is enriched with respect to the solar values, while N/H and O/H appear not to be enhanced, or they present a slight deficiency. It is interesting to notice that the O/H abundance is similar to the one of the IS, but N/H is lower by a factor $\sim$8. It was not possible to estimate the neon abundance here because these lines were not bright enough in the integrated spectra.

We inferred the gas distribution in this shell studying data from the maps in the X$-$ray zone, because, although in our scheme the X$-$ray pointing does not target the OS, it gives us other interesting information. Remembering the statistical frequency velocity distributions in Fig. \ref{fig:histograms}, it can be seen that the overall range in radial velocity covered by [OIII]$\lambda$5007\AA{} and [NII]$\lambda$6584\AA{} emissions is similar ($\sim$80 km~s$^{-1}$), but the [OIII]$\lambda$5007\AA{} velocity distribution shows a single peak for the two defined zones (A and B), whereas the [NII]$\lambda$6584\AA{} emission presents a bi$-$modal distribution in \emph{Zone B} and a redshifted peak in \emph{Zone A}. This behaviour can indicate that the O$^{2+}$ appears to be distributed, forming a rather filled expanding bubble. \\

Analysing the kinematics, abundances, and physical properties, we propose an origin for this shell that could explain why it is only observed in the direction of the minor axis of the ellipse defined above. We argue that ejections of material in the RSG stage were non$-$spherical \citep{2010ASPC..425..247H}, forming an ellipsoidal RSG shell denser towards its major axis. When the WR wind collides with the RSG shell, along the minor axis, the collision does not find strong resistance, fragmenting it and breaking up the MS$-$surrounded bubble (as predicted in GS96 models). The shock then propagates in a tenuous medium where the high temperature favours the [OIII]$\lambda$5007\AA{} emission seen and where a lower cooling rate could prevent the material from the WR stage to be observed in the optical. This idea is basically consistent with previous findings of \citet{2000AJ....120.2670G}.

			    \subsection{Surrounding skin (SS)\label{SS}}
Finally, we propose the existence of an external and faint shell around the whole nebula, the \textit{surrounding skin}. This shell is barely noticeable after close inspection of the INT$-$WFC images, but analysis of the integrated spectra E2 and MB3 shows that a faint emission exists beyond the outer shell. Although the zones where both spectra were extracted do not present any apparent emission in Fig. \ref{fig:integratedspectra}, non negligible emission$-$line fluxes were measured on the spectra.

Properties obtained from these two zones reveal that they belong to a different region than the rest of integrated spectra. Both appear to have similar electron temperatures  and densities, and we find here the lowest N/H abundance (lower than solar). Neither helium nor neon lines were detected in these integrated spectra. The  N/O and O/H in these regions are in the range of the expected values of the ISM at $R_{\mathrm{G}}=10$ kp (NGC\,6888 galactocentric radius) considering the chemical abundance gradients. Also, I(H${\beta}$) has its lowest intensity in E2 (except for the bullet, as explained below). We checked that the emission in this shell is intense enough not to be confused with general diffuse ISM emission, but it is not strong enough to belong to the OS. The derived oxygen and nitrogen abundances appear consistent with the ISM expected values at this galactocentric distance.  \\

Following GS96, we think that this skin should represent the early interaction between the bubble created by winds from the MS star with the circumstellar medium.

			    \subsection{The bullet\label{bullet}}
As we explained in Sect. \ref{intro}, observations in the pointing called Bullet were made to shed light on the nature of a dark object near the centre of the nebula. Measures in this zone (integrated B1) indicate that it is the region where the H${\beta}$ flux (and other emission lines) is lowest, but still not negligible. In this region, abundances of N and O are slightly lower than for the solar values of \citet{2009ARA&A..47..481A}, but  in the same range as the outer shell. The physical properties inferred (T$_{\mathrm{e}}$ and n$_{\mathrm{e}}$) are not remarkable in relation to other regions of the nebula.

\begin{figure}
		  \centering
		 \includegraphics[width=9cm]{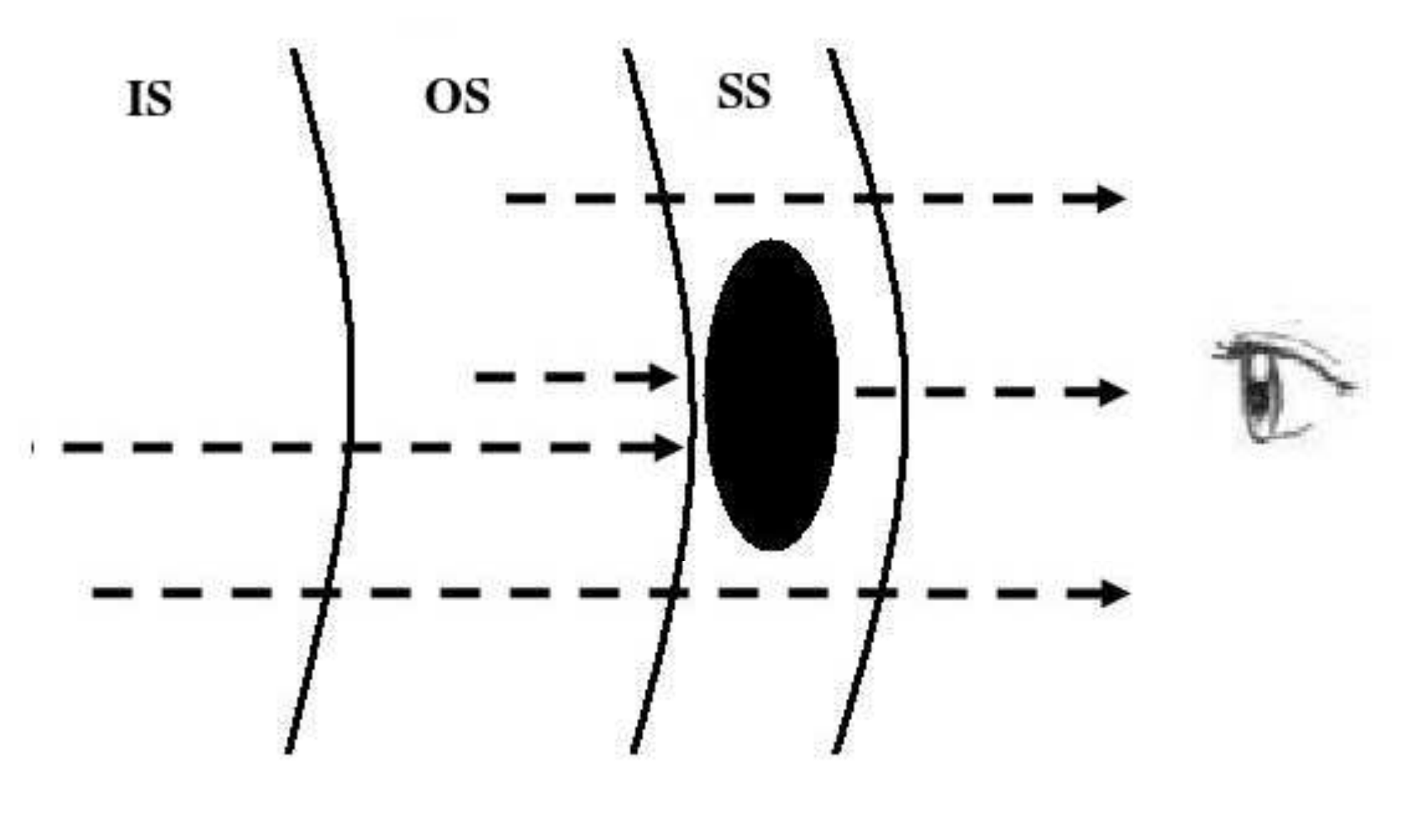}
\caption{Sketch of the effect of the \emph{bullet} on the nebular light in our line of sight. If the bullet is placed in the surrounding skin, it would block the emission from the IS and OS, but not from the edge of the SS.}
      \label{fig:bullet}
 		\end{figure}

Attending to the measured intensities, we propose that this \emph{object} is not behind the nebula because in that case all the nebular emission would be measured in front of it, so it is probably located between the main body of NGC\,6888 and us, but not out of the nebula, because in that case all the emission lines would be blocked. With this analysis and considering the physical properties of B1, we suggest that the bullet is placed in the surrounding skin defined above, and it should only be blocking light coming from the internal shells (see Fig. \ref{fig:bullet}). \\

The observed bubble can also be explained using the GS96 scheme. Maybe such a clump formed in the early collision between MS winds and ISM, or perhaps as an accumulation of molecules and dust of the very cool RSG wind, later impelled by the WR shock. In any case, new observations in other wavelengths, such as radio and infrared, would be necessary to understand the nature and composition of this object.

%________________________________________________________________

\section{Summary and conclusions \label{conclusions}}
\renewcommand {\labelenumi} {\arabic {enumi}$)$}
\renewcommand {\labelenumii} {$\bullet$}
In this article, we have presented an exhaustive study of the wind$-$blown bubble NGC\,6888 with integral field spectroscopic observations carried out with PPAK. The most relevant results presented in this paper can be summarized as follows.
\begin{enumerate}

\item We obtained interpolated images of four regions of the nebula in several wavelengths. We observe two different patterns for the low and high ionization ions. Some features are only observed in the [OIII]$\lambda5007$\AA{} emission line, indicating clear differences in the excitation conditions. \\

\item In the X$-$ray emitting zone of the nebula, a dithering scheme was adopted in the observations. This allowed us to generate cubes to carry out a thorough bi$-$dimensional analysis of this region:\\
\begin{enumerate}
\item We obtained a map for the reddening coefficient, c(H${\beta}$), finding a clear non$-$uniform structure, rather filamentary with a mean value of $\sim$0.57.
\item We derived the electron density map, n$_{\mathrm{e}}$, showing a distribution with a peak density ($\sim$400 cm$^{-3}$) spatially coincident with the zone with the highest emission$-$line fluxes in the nebula.
\item We generated radial velocity maps. The velocity field derived from the H${\alpha}$ emission line is not spatially homogeneous and is moving faster in its southwest part.
\item We studied the ionization conditions plotting pixels from intensity line ratios maps in three diagnostic diagrams. From these, we generated a map defining two main spatially well delimited zones: \emph{Zone A} showing correlations between lines as could be expected from ionization structure, given the different ionization degree of the ions, and where the maximum emission in low excitations species was found; and \emph{Zone B} where the relations between the line ratios show significant scatter in diagnostic diagrams that is located in the area corresponding to the peak of the [OIII]$\lambda5007$\AA{} emission.
\item We compared the statistical frequency distributions of the radial velocity for different lines. The [OIII]$\lambda5007$\AA{} histogram shows a single and centred peak in both zones (A and B). The [NII]$\lambda6584$\AA{} presents one redshifted peak in \emph{Zone A} and a bimodal distribution in \emph{Zone B}. This result could be explained as the two components of an expanding and broken shell.
\item We evaluated the possible presence of shocks to explain the observed behaviours in \emph{Zone B}, while resorting to the models presented by \citet{2008ApJS..178...20A}. We found that our data of \emph{Zone B} can be represented well by models in which shock and a precursor are involved, assuming $n=1000~cm^{-3}$, twice solar abundances, and shock velocities from 250 to 400 km~s$^{-1}$.\\
\end{enumerate}

\item We generated nine integrated spectra extracted from the four pointings of the nebula. We measured line intensities to obtain physical parameters and chemical abundances:\\
\begin{enumerate}
\item We derived electron temperatures, T$_{\mathrm{e}}$, using the $R_{\mathrm{N2}}=[\mathrm{NII}]\lambda\lambda 6548,6584/[\mathrm{NII}]\lambda 5755$ ratio, obtaining values ranging from $\sim$7700~K to $\sim$10\,200~K. From the [SII]$\lambda\lambda$6717/6731 line ratios, electron densities n$_{\mathrm{e}}$, were calculated by deriving differences of up to 350 cm$^{-3}$ between the studied zones.
\item We inferred that nearly all the zones present a slightly underabundance of oxygen with respect to the solar abundances. We derived a nitrogen abundance similar to the solar value in  five regions; however, in other zones N/H appears enhanced up to a factor of 6 or even 8. Helium abundance presents an enrichment in most of the integrated zones, too.
\item We compared selected emission$-$line ratios of the integrated spectra with predictions from photoionization models on diagnostic diagrams. The values obtained with these models for $\log \mathrm{(N/O)}$ agree with the abundance derived. \\
\end{enumerate}

\item We provided a scenario for the evolution of the central WN6 star to explain the features observed in NGC\,6888. The proposed scheme consists of a shell structure, but allows irregular and/or broken shells with different physical, kinematical, and chemical properties:\\
\begin{enumerate}
\item An inner, elliptical and broken shell (IS), with a strong overabundance in N/H and slightly deficient in O/H. We argue that the IS material corresponds to the RSG and WR shells that were shocked and expand in a low$-$density medium.
\item An outer and spherical shell (OS) that is weaker in all the observed emission lines except in [OIII]$\lambda$5007\AA{}. The helium appears enriched here, while N/H and O/H do not appear enhanced or do not present any slight deficiency. We argue that this shell is the early MS bubble broken up as a consequence of the collision between RSG and WR shells.
\item We propose the existence of an external and faint shell around the whole nebula, the \textit{surrounding skin} (SS). Measured fluxes in this regions are the lowest but not negligible. We think that this $“skin”$ should represent the early interaction between the bubble created by winds from the MS star with the interstellar medium. The abundances inferred for this component appear to be typical of the local ISM.
\item We checked that the dark region (\emph{bullet}) observed in the centre of NGC\,6888 appears to be associated with the nebula and is probably located in its most external shell. Observations in other wavelengths, such as radio and infrared, would be necessary to fully understand the nature of this object, but we speculate that it could be a dense clump impelled by the WR winds crossing the shells like a bullet.\\
\end{enumerate}

\end{enumerate}

%________________________________________________________________

\begin{acknowledgements}
This work is supported by the Spanish Ministry of Science and Innovation (MICINN) under the grant BES$-$2008$-$008120. This work has been partially funded by the projects: AYA2007$-$67965$-$C01 and AYA2010$-$21887$-$C04$-$01 of the Spanish PNAYA;  CSD2006 $-$ 00070 \textquotedblleft 1st Science with GTC\textquotedblright  \,from the CONSOLIDER 2010 programme of the Spanish MICINN and  TIC114 of the Junta de Andaluc\'ia. SFS thanks the {\it Viabilidad, Diseno, Acceso y Mejora} funding programme, ICTS$-$2009$-$10, and the {\it Plan Nacional de Investigacion y Desarrollo} funding programme, AYA2010$-$22111$-$C03$-$03, of the Spanish Ministerio de Ciencia e Innovaci\'on, for the support given to this project.
We thank Mercedes Moll\'a for her useful help with the models of the WR stellar atmospheres.
\end{acknowledgements}

%________________________________________________________________

%________________________________________________________________

\begin{landscape}
\begin{table}%[h!]
		 \caption{Measured lines in all the integrated spectra. The intensity values of each emission line (I($\lambda$)/I(H${\beta}$)) are normalized to $F(H{\beta})=100$ and reddening corrected.}    % title of Table
\label{table:all_lines}     % is used to refer this table in the text
		\centering                          % used for centering table
		\begin{tabular}{l c c c c c c c c c c c }
		\hline
        &&&\multicolumn{9}{c}{I($\lambda$)/I(H${\beta}$)} \\
       \cline{4-12}
Line & $\lambda~(\AA{})$ & $f(\lambda)^{a}$ &  X1$^{b}$ & X2$^{b}$ & E1$^{b}$ & E2$^{b}$ & MB1 & MB2 & MB3 & B1 & B2 \\
		\hline
		\hline
         \\
		{[}OII]   & 3728 & 0.322 & 125.7 $\pm$ 13.1 & 181.0 $\pm$ 3.5 & 281.4 $\pm$ 36.1 & 217.0 $\pm$ 19.4 & 178.5 $\pm$ 13.2 & 274.1 $\pm$ 31.9 & 279.6 $\pm$ 35.1 & 234.4 $\pm$ 38.6 & 112.8 $\pm$ 12.9 \\
		H9 & 3835 & 0.298 & 10.0 $\pm$ 1.5 & 9.0 $\pm$ 3.2 $\dagger$ & 25.0 $\pm$ 7.1 $\dagger$ & ... & 10.8 $\pm$ 1.7 $\dagger$ & ... & ... & ...  & 9.6 $\pm$ 2.3 \\
		{[}NeIII]  & 3868 & 0.291 & 10.8 $\pm$ 2.8 & 31.1 $\pm$ 9.3 & 43.4 $\pm$ 16.3 $\dagger$ & ... & 4.1 $\pm$ 1.1 & 29.8 $\pm$ 8.4 $\dagger$ & 51.8 $\pm$ 15.1 $\dagger$ &...  & 16.1 $\pm$ 1.5 \\
		H8+HeI  & 3889 & 0.286 & 29.7 $\pm$ 2.4 & 27.3 $\pm$ 3.5 & 6.9 $\pm$ 1.8 $\dagger$ & ... & 23.7 $\pm$ 2.3 & ... & ...& ... & 30.4 $\pm$ 0.8 \\
		H7 & 3970& 0.266 & 16.3 $\pm$ 0.8 & 16.0 $\pm$ 2.5 $\dagger$ & ... & ... & 16.5 $\pm$ 3.9 & ... & ... & ...  & 21.0 $\pm$ 3.8 \\
		HeI+[NII]  & 4026 & 0.251 & 4.4 $\pm$ 0.4 & 3.5 $\pm$ 0.7 $\dagger$ & ... & ... & ...  & ... & ... & ... & ... \\
		H${\delta}$  & 4101 & 0.229 & 29.2 $\pm$ 0.1 & 31.3 $\pm$ 0.4 & 27.6 $\pm$ 3.6 & 26.8 $\pm$ 4.0 & 28.1 $\pm$ 1.5 & 29.3 $\pm$ 3.2 & 26.8 $\pm$ 4.8 & 41.5 $\pm$ 7.4 $\dagger$ & 30.0 $\pm$ 1.8 \\
		H${\gamma}$ & 4340 & 0.156 & 52.6 $\pm$ 0.2 & 57.8 $\pm$ 0.8 & 49.4 $\pm$ 1.2 & 47.9 $\pm$ 4.5 & 53.0 $\pm$ 1.6 & 54.4 $\pm$ 11.5 & 55.0 $\pm$ 12.7 & 57.0 $\pm$ 14.3 $\dagger$ & 55.8 $\pm$ 1.9 \\
		{[}OIII] & 4363 & 0.149 & 3.5 $\pm$ 0.3 $\dagger$ & 18.8 $\pm$ 2.7 $\dagger$ & 27.4 $\pm$ 14.8 $\dagger$ & ... & ... & ...  & ... & ... & ... \\
		HeI   & 4471 & 0.115 & 10.1 $\pm$ 0.7 & 9.5 $\pm$ 0.6 & 5.5 $\pm$ 1.2 & ... & 7.7 $\pm$ 0.8 & ...  & ... & ... & 9.0 $\pm$ 0.8 \\
		{[}OIII]   & 4959 & -0.026 & 45.9 $\pm$ 0.2 & 101.0 $\pm$ 1.0 & 108.2 $\pm$ 2.9 & 39.9 $\pm$ 0.1 & 24.4 $\pm$ 0.6 & 111.5 $\pm$ 7.0 & 51.9 $\pm$ 4.7 & 39.3 $\pm$ 2.3 & 81.2 $\pm$ 1.0 \\
		{[}OIII]   & 5007 & -0.038 & 138.1 $\pm$ 2.7 & 299.1 $\pm$ 0.4 & 328.9 $\pm$ 11.7 & 119.9 $\pm$ 1.6 & 74.1 $\pm$ 1.0 & 335.5 $\pm$ 8.7 & 148.5 $\pm$ 9.3 & 110.4 $\pm$ 13.3 & 242.7 $\pm$ 1.3 \\
		HeI   & 5015 & -0.040 & 3.7 $\pm$ 0.3 & 1.2 $\pm$ 0.1 & ... & ... & ... & ... & ...  & ... & ...   \\
		{[}NI]  & 5200 & -0.083 & ...   & ... & 12.4 $\pm$ 2.6 & ... & 4.1 $\pm$ 0.3 & 8.5 $\pm$ 1.7 $\dagger$ & 12.1 $\pm$ 3.0 & ... & 1.9 $\pm$ 0.3 \\
		{[}NII]  & 5755 & -0.185 & 3.0 $\pm$ 0.1 & ... & ... & ... & 3.3 $\pm$ 0.2 & ... & ... & ... & 1.2 $\pm$ 0.3 \\
		HeI  & 5876 & -0.203 & 28.2 $\pm$ 1.4 & 31.6 $\pm$ 3.1 & 20.4 $\pm$ 5.9 & ... & 22.6 $\pm$ 1.0 & 19.9 $\pm$ 3.1 & ... & ... & 23.5 $\pm$ 0.9 \\
		{[}SIII]   & 6312 & -0.264 & 1.3 $\pm$ 0.1 $\dagger$ & ...  & ...  & ... & ...   & ... & ... & ... & ... \\
		{[}NII]    & 6548 & -0.296 & 135.8 $\pm$ 2.8 & 90.9 $\pm$ 2.6 & 29.2 $\pm$ 8.5 & 25.0 $\pm$ 6.7 & 199.2 $\pm$ 10.5 & 44.3 $\pm$ 7.2 & 27.2 $\pm$ 6.0 & 38.9 $\pm$ 6.2 & 48.3 $\pm$ 1.8 \\
		H${\alpha}$  & 6563 & -0.298 & 324.4 $\pm$ 0.3 & 351.6 $\pm$ 0.3 & 305.8 $\pm$ 16.0 & 295.0 $\pm$ 9.7 & 317.1 $\pm$ 16.3 & 330.8 $\pm$ 22.2 & 313.6 $\pm$ 20.4 & 293.2 $\pm$ 43.3 & 285.1 $\pm$ 4.2 \\
		{[}NII]    & 6584 & -0.300 & 418.1 $\pm$ 2.2 & 274.3 $\pm$ 6.2 & 83.5 $\pm$ 11.9 & 77.5 $\pm$ 9.0 & 581.4 $\pm$ 29.4 & 133.1 $\pm$ 11.3 & 76.3 $\pm$ 5.6 & 111.0 $\pm$ 17.0 & 138.4 $\pm$ 2.8 \\
		HeI   & 6678 & -0.313 & 7.0 $\pm$ 0.1 & 6.9 $\pm$ 0.3 & ... & ... & 6.0 $\pm$ 0.4 & ... & ... & 3.2 $\pm$ 0.6 $\dagger$ & 6.2 $\pm$ 0.2 \\
		{[}SII]    & 6717 & -0.318 & 16.9 $\pm$ 0.1 & 25.3 $\pm$ 0.5 & 43.2 $\pm$ 3.0 & 46.0 $\pm$ 2.9 & 34.0 $\pm$ 2.1 & 31.5 $\pm$ 2.8 & 36.5 $\pm$ 4.0 & 42.1 $\pm$ 6.9 & 9.8 $\pm$ 0.1 \\
		{[}SII]    & 6731 & -0.320 & 13.9 $\pm$ 0.3 & 19.1 $\pm$ 0.1 & 28.5 $\pm$ 2.4 & 29.7 $\pm$ 2.4 & 33.2 $\pm$ 2.1 & 20.5 $\pm$ 2.5 & 23.8 $\pm$ 3.3 & 31.7 $\pm$ 5.6 & 7.4 $\pm$ 0.2 \\
           \\
       F(H${\beta}$)$^{c}$ & & & 195.7  $\pm$ 3.2   &  43.0 $\pm$ 0.8  & 3.4  $\pm$ 0.1   &   3.2  $\pm$ 0.2 &   39.1 $\pm$ 0.3    &  5.5  $\pm$ 0.2   &  3.5  $\pm$ 0.2 & 2.7  $\pm$ 0.2   &  31.3 $\pm$ 0.2   \\
       c(H${\beta}$)   &   &   &  0.75 $\pm$ 0.01 &  0.77 $\pm$ 0.01 &  0.67 $\pm$ 0.04  &   0.66 $\pm$ 0.02  &  0.72 $\pm$ 0.07 &  0.60 $\pm$ 0.09 &   0.69 $\pm$ 0.09 & 0.73 $\pm$ 0.21  & 0.71 $\pm$ 0.02 \\
         \\
		\hline
		\end{tabular}
        \begin{list}{}{}
                  \item {$^{a}$} Adopted reddening curve. Obtained using \citet{1989ApJ...345..245C} extintion law with $R_{\mathrm{V}}=3.1$.  \\
                  \item {$^{b}$}  In these regions intensities are the mean weighted by errors of the different emission lines measured from high and low resolution gratings.\\
                  \item {$^{c}$} H${\beta}$ fluxes in units of $x10^{-15}~erg~cm^{-2}~s^{-1}$ (not corrected for extinction).\\
                  \item {$\dagger$} Lines with bad measure. These lines are not used in determining physical parameters or abundances. \\
		\end{list}
		\end{table}
\end{landscape}

\begin{landscape}
\begin{table}%[h!]
		 \caption{Electron densities (cm$^{-3}$) , temperatures ($\times 10^{-4}$~K), ionic abundances, and total chemical abundances for the nine integrated spectra.}    % title of Table
		\label{table:paramyabun}     % is used to refer this table in the text
		\centering                          % used for centering table
		\begin{tabular}{l c c c c c c c c c c}
		\hline
		& X1 & X2&  E1 & E2 & MB1 & MB2 & MB3 & B1 & B2 & Solar$^{*}$\\
		\hline
		\hline
        \\
		n$_{\mathrm{e}}$([SII]) & 179$\pm$ 18 & 106$\pm$ 16 & $<$100  & $<$100  & 363$\pm$ 116 & $<$100 & $<$100  & 105 $\pm$ 179 & 108 $\pm$ 26 &... \\
		\\
		T$_{\mathrm{e}}$([NII]) & 8101 $\pm$ 85 &9747$\pm$ 108 $\dagger$  &10171$\pm$ 729 $\dagger$ & 8565$\pm$ 220 $\dagger$  &7750$\pm$ 72 &10199 $\pm$ 675 $\dagger$  & 8956$\pm$ 385 $\dagger$  & 8552 $\pm$ 412 $\dagger$  & 8537 $\pm$ 527 &...  \\
		T$_{\mathrm{e}}$([OIII])$_{E}$  & 6345 $\pm$ 98 &8416 $\pm$ 151 &9019 $\pm$ 1079 & 6888 $\pm$ 268 &5952 $\pm$ 79 &9061 $\pm$ 1003 & 7371 $\pm$ 491 & 6873 $\pm$ 500 & 6855 $\pm$ 639 &... \\
		T$_{\mathrm{e}}$([OII])$_{E}$ & 7165$\pm$ 185 &9016$\pm$ 224 &10702$\pm$ 1426 & 9805$\pm$ 7308 &6784 $\pm$ 386 &11337 $\pm$ 6066 & 9382$\pm$ 6195 & 7737 $\pm$ 1138 & 7715 $\pm$ 1055 &... \\
		T$_{\mathrm{e}}$([SII])$_{E}$  & 6288$\pm$ 131 &7602$\pm$ 159 &8798$\pm$ 1012 & 8161$\pm$ 5189 &6017$\pm$ 274 &9249 $\pm$ 4307 & 7861$\pm$ 4398 & 6693 $\pm$ 808 & 6679 $\pm$ 749 &... \\
		\\	
		12+log(O$^{+}$/H$^{+}$) & 8.09 $\pm$ 0.03 & 7.89 $\pm$ 0.02 & 7.99 $\pm$ 0.14 & 8.22 $\pm$ 0.07 & 8.40 $\pm$ 0.04 & 7.98 $\pm$ 0.13 & 8.24 $\pm$ 0.10 & 8.27 $\pm$ 0.13 & 7.96 $\pm$ 0.14  &...  \\
		12+log(O$^{2+}$/H$^{+}$)& 8.49 $\pm$ 0.03 & 8.27 $\pm$ 0.03 & 8.20 $\pm$ 0.19 & 8.25 $\pm$ 0.08 & 8.36 $\pm$ 0.03 & 8.20 $\pm$ 0.18 & 8.22 $\pm$ 0.13 & 8.23 $\pm$ 0.15 & 8.57 $\pm$ 0.19   &... \\
		12+log(S$^{+}$/H$^{+}$) & 6.12 $\pm$ 0.01 & 6.04 $\pm$ 0.01 & 6.19 $\pm$ 0.08 & 6.43 $\pm$ 0.04 & 6.54 $\pm$ 0.02 & 6.04 $\pm$ 0.08 & 6.27 $\pm$ 0.06 & 6.43 $\pm$ 0.08 & 5.80 $\pm$ 0.08  &...  \\
		12+log(N$^{+}$/H$^{+}$) & 8.17 $\pm$ 0.02 & 7.76 $\pm$ 0.02 & 7.20 $\pm$ 0.10 & 7.37 $\pm$ 0.06 & 8.38 $\pm$ 0.02 & 7.39 $\pm$ 0.08 & 7.31 $\pm$ 0.06 & 7.53 $\pm$ 0.08 & 7.63 $\pm$ 0.08  & ... \\
		12+log(Ne$^{2+}$/H$^{+}$) & 8.07 $\pm$ 0.12 & 7.87 $\pm$ 0.14 & ... & ... & 7.83 $\pm$ 0.13 & ... & ... & ... & 8.05 $\pm$ 0.24  &...  \\
		(He$^{+}\lambda$4471/H$^{+}$) & 0.20 $\pm$ 0.01 & 0.19 $\pm$ 0.01 & 0.11 $\pm$ 0.02 & ... & 0.16 $\pm$ 0.02 & ... & ... & ... & 0.17 $\pm$ 0.02  &...  \\
		(He$^{+}\lambda$5875/H$^{+}$) & 0.20 $\pm$ 0.01 & 0.22 $\pm$ 0.02 & 0.15 $\pm$ 0.04 & ... & 0.16 $\pm$ 0.01 & 0.14 $\pm$ 0.02 & ... & ... & 0.16 $\pm$ 0.01  & ... \\
		(He$^{+}\lambda$6678/H$^{+}$) & 0.17 $\pm$ 0.01 & 0.17 $\pm$ 0.01 & ...  & ... & 0.14 $\pm$ 0.01 &  ... & ... & ... & 0.15 $\pm$ 0.01  &...  \\
		\\
		ICF(Ne) & 1.09 $\pm$ 0.01 & 1.10 $\pm$ 0.01 & ... & ... & 1.18 $\pm$ 0.02 & ... & ... & ... & 1.08 $\pm$ 0.01  &... \\
		\\	
		12+log(O/H) & 8.64 $\pm$ 0.03 & 8.42 $\pm$ 0.02 & 8.41 $\pm$ 0.13 & 8.54 $\pm$ 0.05 & 8.68 $\pm$ 0.03 & 8.40 $\pm$ 0.12 & 8.53 $\pm$ 0.08 & 8.55 $\pm$ 0.10 & 8.66 $\pm$ 0.16 & 8.69 $\pm$ 0.05 \\
		12+log(N/H) & 8.72 $\pm$ 0.04 & 8.29 $\pm$ 0.04 & 7.61 $\pm$ 0.22 & 7.68 $\pm$ 0.10 & 8.67 $\pm$ 0.05 & 7.82 $\pm$ 0.20 & 7.60 $\pm$ 0.15 & 7.82 $\pm$ 0.18 & 8.34 $\pm$ 0.23 & 7.83 $\pm$ 0.05\\
		12+log(Ne/H) & 8.11 $\pm$ 0.12 & 7.91 $\pm$ 0.14 & ... & ... & 7.90 $\pm$ 0.13 & ... & ... & ... & 8.08 $\pm$ 0.24  &7.93 $\pm$ 0.10\\
		log(N/O) & 0.08 $\pm$ 0.04 & -0.13 $\pm$ 0.03 & -0.79 $\pm$ 0.17 & -0.85 $\pm$ 0.09 & -0.01 $\pm$ 0.05 & -0.58 $\pm$ 0.16 & -0.93 $\pm$ 0.12 & -0.74 $\pm$ 0.15 & -0.32 $\pm$ 0.16  & -0.86 $\pm$ 0.07\\
		log(Ne/O) & -0.52 $\pm$ 0.11 & -0.52 $\pm$ 0.13 & ... & ... & -0.79 $\pm$ 0.12 & ... & ... & ... & -0.58 $\pm$ 0.07  & -0.76 $\pm$ 0.11\\
		y & 0.17 $\pm$ 0.01 & 0.18 $\pm$ 0.01 & 0.12 $\pm$ 0.01 & ... & 0.15 $\pm$ 0.01 & 0.14 $\pm$ 0.02 & ... & ... & 0.16 $\pm$ 0.01 & 0.09 $\pm$ 0.01\\		
		\\	
		\hline
		\end{tabular}
        \begin{list}{}{}
                  \item {$_{E}$} Temperatures derived from other $T_{\mathrm{e}}$. See text for details  \\
                  \item {$\dagger$} Estimated from the empirical flux$-$to$-$flux relation proposed by \citet{2007MNRAS.375..685P}.\\
                  \item {$^{*}$} Solar values for the total abundances from \citet{2009ARA&A..47..481A}.\\
		\end{list}
		\end{table}
\end{landscape}

% end of the main text

\clearpage \onecolumn
\end{document}